# Modern theory of hydraulic fracture modeling
# with using explicit and implicit schemes

## A. M. Linkov
Rzeszow University of Technology, Poland
e-mail: linkoval@prz.edu.pl

**Abstract**. The paper presents a summary of the novel results, obtained on the basis of the modified theory of hydraulic fractures (HF) developed by the author. The theory underlines significance of the speed equation (SE) as a necessary component of the system of equations to have it complete and well-posed. When applied to numerical simulation of HF, the theory reveals three distinct issues important for efficient modeling. These are: (i) modeling the central part of a HF; (ii) modeling the near-front zone; and (iii) tracing changes in the shape of a fracture contour.

Modeling the *central part* leads to a stiff system of ODE in time, what strongly complicates its integration. For *explicit* schemes, it requires *small time steps* to meet the CFL condition of stability. For *implicit* schemes, it requires *proper preconditioners* to provide convergence with non-excessive number of iterations. The gains and flaws of the two strategies are discussed in detail. It is noted that a rough spatial mesh may be used in the central part.

Modeling the *near-front* zone reveals the vital role of the *speed equation* (SE) for HF modeling by any method. Its asymptotic analysis has resulted in the fundamental concept of the *universal asymptotic umbrella* (UAU). For a smooth part of the front in a homogeneous part of a medium, the UAU universally connects the major characteristics of the near-front zone (the opening, the propagation speed, the distance to the front), what offers remarkable analytic and computational gains. Firstly, it gives the absolute value of the propagation speed. Secondly, it provides the fluxes on the internal sides of tip elements needed to meet the discretized continuity equation. Furthermore, for the near-front zone, it is established that a notable part of the zone adjacent to the front propagates virtually as a *simple wave*. This implies that the CFL condition of stability for this zone is much less restrictive than for the central part of the fracture. For a homogeneous medium, the size of the zone propagating as a simple wave is large enough to employ a rough mesh with some 5-10 elements along a typical half-length of the homogeneous region. Joined with the similar statement for the central part, this leads to the key conclusion that a rough spatial mesh may serve when modeling HF in a homogeneous region. For a HF, propagating through local stress barriers, the CFL condition may become even less restrictive due to drop of the propagation speed. This suggests that fine spatial meshes and hybrid schemes may be of use to model HF propagation through barriers. Of essence is also that the wave-like propagation of the near front zone makes preferable upwind schemes of time stepping (especially, when employing *explicit* integration). On whole, the analysis implies that *explicit time stepping may be complementing, competitive and even superior* over implicit integration.

Tracing *changes in the front shape* appears merely in 3D problems when the contour changes its form. Since the speed vector is represented in the UAU only by its magnitude, there is need in *reconstruction the front and the normal to it* on some time steps. It is shown that the reconstruction of smooth parts of a front, propagating in homogeneous parts of a medium, is reduced to building the envelope to the distance-circles defined by the UAU. The problem, being essentially geometrical, it may be solved separately by various methods, including fast marching, level set and the simplest string/marker methods. In 1D cases, it does not arise at all.

Quantitative estimations and numerical examples illustrate the theoretical conclusions.

## 1. Introduction

Hydraulic fracturing (HF) is a process of fracture propagation induced by the pressure of an injected fluid. It is widely used in various industry applications including stimulation of oil, gas and heat recovery, $CO_2$ sequestration, waste disposal, excavation of hard rock, roof control and preconditioning in mines, etc. It also occurs in natural conditions when magma breaks through the crust to the earth surface. In view of limited and uncertain data on the rock structure and properties at the injection depth, the mathematical analysis and numerical simulation are of prime importance to increase understanding and to enhance efficiency of HF. Their significance has been explained in detail in the review by Mack and Warpinski (2000). Since the pioneering works by Khristianovich and Zheltov (1955), Carter (1957), Perkins and Kern (1961), Geertsma and de Klerk (1969), Howard and Fast (1970), Spence and Sharp (1985), numerous studies have been performed (e.g. [Kemp 1990; Descroches J., et al 1994, Lenoach 1995, Economides and Nolte 2000, Garagash and Detournay 2000, Adachi and Detournay 2002, Savitski and Detournay 2002, Peirce and Siebrits 2005, Peirce 2006, Garagash 2006, Zubkov et al. 2006, Adachi et al. 2007, Peirce and Detournay 2008, Garagash, Detournay and Adachi 2011, Gordeliy and Detournay 2011, Linkov 2011, 2012, Mishuris et al. 2012, Gordeliy and Peirce 2013, Lecampion et al. 2013, Napier and Detournay 2013, Wrobel and Mishuris 2015, Dontsov and Peirce 2015, Linkov 2015, Golovin et al. 2015, Peirce 2015, 2016, Cherny et al. 2016, Lecampion, Bunger and Zhang 2018, Cao, Hassein and Schreffler 2018]).

The summary of theoretical results to 2007 may be found in the review by Adachi et al. (2007). A review of the state-of-art to 2015 is given in the paper by Peirce [Peirce 2016], containing also original results of the author. The latest review by Lecampion et al. [Lecampion, Bunger and Hi 2018] on whole follows the reviews [Adachi et al. 2007, Peirce 2016] complementing them with additional references. The paper [Peirce 2016] contains the commonly used system of equations and outlines the importance of proper accounting for asymptotic behavior of quantities near the fracture contour, studied in many papers (e.g. [Descroches et al. 1994; Garagash and Detournay 2000, Garagash 2006, Garagash, Detournay and Adachi 2011; Peirce 2015]) starting from the paper by Spence and Sharp [Spence and Sharp 1985] for the viscosity dominated regime and the paper of Lenoach [Lenoach 1995] for the leak-off dominated regime. The paper [Peirce 2016] contains also detailed description of the most advanced method of modeling HF suggested by Peirce and Detournay [Peirce and Detournay 2008]. The method, named the implicit level set algorithm (ILSA), is presently superior over other methods implemented in HF simulators. This has been established in the special comparative study [Lecampion et al. 2013] (see also [Lecampion, Bunger and Zhang 2018]). The advantages of the ILSA have been evidently demonstrated by impressive examples given in the papers [Gordeliy and Peirce 2013, Peirce 2015, 2016]. Therefore, in the following discussion, the ILSA may serve us as a sample to match our results. Since the essence of the method is in its implicit time stepping, we, for brevity, sometimes will call it simply *implicit method*. The method employs the *backward Euler* scheme, which avoids highly restrictive Courant-Friedrich-Levy (CFL) condition on the size of a time step. The latter may be even as large as the time needed to the fracture contour to propagate a distance of a spatial mesh size.

The price for this advantage, as usual for implicit schemes, is the need to solve a strongly non-linear system within a time step. The latter, in its turn, requires, firstly, finding a proper fixed point operator (in an external cycle) to overcome non-linearity and, secondly, developing a proper preconditioner to solve a linearized algebraic system (in an internal cycle). Various fixed point iterations and methods to effectively solve a linearized system are discussed in [Peirce 2016]. Specifically, the original ILSA [Peirce and Detournay 2008; Peirce 2015, 2016] employed the preconditioners suggested in [Peirce and Siebrits 2005, Peirce 2006] for HF driven by a Newtonian fluid. An extension to non-Newtonian fluids requires further work. The shortcomings of the implicit schemes impel looking for complementary means to simulate HF, such as *explicit* rather than implicit time stepping.

Meanwhile, to the date, the cited most comprehensive reviews [Adachi et al 2007; Peirce, 2016; Lecampion, Bunger and Hi 2018)] have focused on extremely high stiffness of coupled spatially discretized elasticity-fluid equations. They have denounced the option of using explicit schemes as "a poor choice" [Lecampion, Bunger and Hi 2018, p. 72], "prohibitively expensive" [Peirce and Siebrits 2005, p. 1822], "extremely computationally intensive" [Peirce and Detournay 2008, p. 2859], because of "prohibitively small time steps" [Adachi et al. 2007, p. 750], required by the CFL condition. The question arises: *is indeed the time step prohibitively small*?

Turning to *explicit* time stepping, note that it has been systematically employed in papers developing the modified formulation of the HF problem with distinct accounting for the SE in its asymptotic form [Linkov 2011 a, b; 2012, 2013, 2014 a, b; 2016 a, b, c; Mishuris et al. 2012; Linkov and Mishuris 2013]. However, these papers concerned with merely 1D problem for a straight fracture and employed *ad hoc* favorable features of this problem. Namely, (i) for the Perkins-Kern-Nordgren (PKN) [Perkins and Kern 1961; Nordgren 1972], Khristianovich-Geertsma-de Klerk  (KGD) [Khristianovich and Zheltov 1955; Geertsma and de Klerk 1969] models and for the axisymmetric problem, the spatial coordinate can be normalized by the fracture half-length, and (ii) for the KGD and axisymmetric problems, the weakly singular inversion of the hypersingular elasticity operator is available in an analytical from. The first favorable feature, employed in all the cited papers, excludes singularity of the temporal derivative of the opening at the tip, because the derivative is evaluated under constant normalized coordinate, and consequently it identically equals to zero at the fracture tip. The second feature, following for the KGD model from the classical Muskhelishvili's solution [Muskhelishvili 1975], has been commonly employed to obtain the benchmark solution (e.g. [Spence and Sharp 1985, Appendix A; Garagash and Detournay 2000, Adachi and Detournay 2002, Garagash 2006, Wrobel and Mishuris 2015)]). When employing these features or even the first of them, explicit methods like Runge-Kutta, Adams and forward Euler, provide reasonable efficiency [Linkov 2011, 2012; Mishuris et al. 2012, Linkov and Mishuris 2013, Linkov 2016 a, b]. However, the both favorable features are available only in the particular cases of the KGD and axisymmetric problems. For a general case of a 2D fracture propagating in the 3D space, it is of prime significance to investigate explicit schemes without using these favorable features. Below we consider explicit and implicit stepping free of the *ad hoc* simplifications.

An attempt to apply an explicit scheme to the KGD problem without using the *ad hoc* simplifications has been made in the paper [Stepanov and Linkov 2016]. For this model, the time step $\Delta t$, required by the CFL condition, is of order $\Delta t = O(\Delta x^3)$, where $\Delta x$ is the size of a spatial mesh [Peirce and Siebrits 2005; Peirce 2006]. It has appeared that the time step is not prohibitive for computations on a conventional laptop. Although the conclusion referred to the 1D problem with the number of unknowns not exceeding first hundreds, it was quite promising. Estimations and calculations for a planar HF, presented below, confirm that, when using results of the theoretical analysis, explicit integration is not a poor choice.

Mention may be also made on the recent papers, which actually employ implicit schemes in ways different of that of the ILSA. In the paper [Golovin et al. 2015], the 1D problem for a straight crack was solved by the finite element method not using the speed equation and asymptotics. Five time steps, performed by the authors, brought signs of notable front smearing.

The papers [Napier and Detournay 2013; Cherny et al. 2016] contained attempts to trace the out-of-plane propagation of the fracture initiated at a borehole. These papers make a step in extending the results for a curvilinear plane-strain fracture [Zubkov et al. 2006], to the 3D problem. The fracture front was traced by equating the tensile stress intensity factor (SIF) $K_I$ to its critical value $K_{IC}$. The direction of the out-of-plane motion was found from the conditions of zero shear SIFs $K_{II}$ and $K_{III}$. Naturally, the authors focused on iterative satisfying these conditions. Specifically, in the paper [Cherny et al. 2016], the algorithm for toughness dominated regime, included two iterative cycles, external to meet the fracture condition $K_I = K_{IC}$, and internal to adjust the trajectory to the condition $K_{II} = 0$. Simulations of the viscosity dominated regime were also performed with two iterative cycles within a time step. The algorithms, being computationally intensive, rough spatial meshes were employed. In this paper, we present results partly applicable to simulation of the out-of-plane HF propagations. Still, the discussion below is focused on a planar HF.

The paper aims to summarize and extend the modified theoretical rationale of HF and to show that it suggests efficient using both *implicit and explicit* schemes of time stepping. The paper presents an extended version of the Key-Note lecture [Linkov 2018].

The research below shows that the CFL stability condition for explicit integration of the ODE is not prohibitive in the time expense. The theoretical analysis reveals why rough meshes are acceptable for modeling HF. We state that for *rough* meshes, the time expense of explicit schemes, even without accelerations, is on the level of implicit integration. Furthermore, using *local* fine meshes near the front does not lead to computational instability. For *fine* meshes on the *entire* fracture, the favorable feature of the explicit integration, which employs matrix-to-vector multiplications with a single fully-populated matrix evaluated in advance, suffices means (fast Fourier transform, fast multipole methods, easy parallel computations) for drastic reduction of the expense. Using explicit integration avoids the mentioned drawbacks of implicit integration. In particular, numerical experiments confirm that it is of immediate use for tracing fractures driven by a non-Newtonian fluid. From the experiments, it also appears that explicit schemes are very stable under strong perturbation of initial conditions. We conclude that explicit integration of the system, presents a perspective competitive/complementing tool for modeling HF.

## 2. Modified problem formulation

The problem formulation includes (i) equations (fluid and elasticity) for internal points of the fracture and (ii) propagation equations for points at the fluid front and (iii) propagation equations for points at the fracture contour (Fig. 1). Since the equations for internal points contain spatial and temporal derivatives of physical quantities, the problem formulation includes also boundary conditions (BC) and initial conditions (IC).

The first of fluid equations for *internal* points is the mass conservation law, which for incompressible fluids expresses volume conservation. The volume conservation (continuity) equation is

$$\frac{\partial w}{\partial t} = -\mathrm{div}(w\boldsymbol{v}) - q_l - Q \qquad (1)$$

The second of fluid equations is the movement equation of the Poiseuille type

$$\boldsymbol{v} = -\left(\frac{w^{n+1}}{\mu'}\right)^{1/n} |\nabla p_f|^{1/n-1} \nabla p_f \qquad (2)$$

Fig. 1

In (1) and (2), $w$ is the opening, $t$ is the time, $p_f$ is the fluid pressure, $q_l \geq 0$ and $Q \geq 0$ are the terms accounting for, respectively, fluid leak-off and fluid influx into the fracture, $n$ is the fluid behavior index, $\mu' = 2[2(2n+1)/n]^n M$, $M$ is its consistency index. For a Newtonian fluid, $n = 1$, $\mu' = 12\mu$ with $\mu = M$ being the dynamic viscosity.

Substitution (2) into (1) joins them as the single scalar *lubrication* equation in scalar quantities $w$ and $p_f$. The elasticity equations serve to connect $w$ and $p_f$. Clearly, by using the definition $\boldsymbol{q} = \boldsymbol{v}w$ of the flux $\boldsymbol{q}$ as the product of the velocity $\boldsymbol{v}$ and the opening $w$, the movement equation (2) may be written as the equation for the flux. It is sufficient to multiply (2) by $w$. We shall not re-write (2) in terms of the flux, because using the primary physical quantity, the *particle velocity*, provides some gains when normalizing variables and it has clear connection with the speed equation discussed below. The option to use the flux, when convenient, remains available after the normalizing.

The general elasticity equations for internal points express the traction vector on the fracture surface via the displacement discontinuities. The general hypersingular equations for blocky elastic media, connecting tractions and displacements on interfaces of structural blocks, are given in [Linkov 1995] (see also [Linkov, Zubkov and Kheib 1997; Linkov 2002]). On *plane* parts of the fracture in a homogeneous media, these equations include a separate scalar equation, which connects merely *normal components* of the traction (pressure) and the displacement discontinuity (fracture opening). In the particular case of a planar fracture with the surface $S$ and contour $L_t$, the equation has the commonly used simplest form (e.g. [Linkov and Mogilevskaya 1986; Lin'kov, Zubkov and Kheib 1997; Adachi et al. 2007; Peirce and Detournay 2008]):

$$-\frac{E'}{8\pi} \int_S \frac{w(\xi)}{R^3} \, dS_\xi = p_f(\boldsymbol{x}) - \sigma_{0n}(\boldsymbol{x}), \tag{3}$$

where $E' = E/(1 - \nu^2)$ is the plane-strain elasticity modulus, $E$ is the Young's modulus, $\nu$ is the Poisson's ratio, $R = \sqrt{(x_1 - \xi_1)^2 + (x_2 - \xi_2)^2}$, $\sigma_{0n}$ is in-situ traction normal to the fracture surface, the coordinates $x_1$ and $x_2$ of the right Cartesian system are located in the fracture plane, the coordinate $x_3$ is orthogonal to them; the normal to the plane has the direction of the $x_3$ axis. To simplify notation, compressive stresses and tractions, as well as the fluid pressure, are assumed positive (thus $\sigma_{0n} > 0$). In the lag-zone between the fluid front and the fracture contour, the pressure $p_f$ is assumed zero or equal to a prescribed evaporation value. If not neglecting the lag, the same pressure is prescribed at the fluid front as a boundary condition. For the fracture contour $L_t$, the condition of zero opening serves as the boundary condition: $w(\boldsymbol{x}_*, t) = 0$. This condition is met identically when looking for a solution of the hypersingular equation (3) on the class of functions equal to zero at the contour $L_t$. This completes the issue of boundary conditions.

It is convenient to use the net-pressure $p(\boldsymbol{x}) = p_f(\boldsymbol{x}) - \sigma_0$, rather than the fluid pressure $p_f$ itself, by subtracting a reference rock pressure $\sigma_0$ from $p_f$. For certainty, $\sigma_0$ may be taken as the closing rock pressure at the depth of a borehole. Then defining the stress contrast as $\Delta\sigma(\boldsymbol{x}) = \sigma_{0n}(\boldsymbol{x}) - \sigma_0$, the elasticity equation may be written in terms of the net-pressure and stress contrast by changing $p_f$ to $p$, and $\sigma_{0n}$ to $\Delta\sigma$. Clearly, since $\sigma_0 = const$, we have $\mathrm{grad}\,p_f = \mathrm{grad}\,p$; hence in equation (2), the net-pressure $p$ may replace the fluid pressure $p_f$, as well. These replacements are assumed in further discussion; from now on, the subscript $f$ is omitted and merely net-pressure and stress contrast are considered in the equations for the internal points of a fracture.

The equations, governing the propagations of (i) fluid front and (ii) fracture contour are, respectively, the *speed equations for* the *fluid* front and the *speed equations for* the *fracture* contour (fracture conditions).

In the conventional formulation (see reviews in [Adachi et al. 2007, Peirce 2016, Lecampion et al. 2018]), the fluid speed equation (SE) is not distinguished explicitly. Sometimes, using the SE is even denounced [Peirce Detournay 2008, p. 2870; Detournay and Peirce 2014, p.153], when authors employ the fluid flux instead of the primary physical quantity, the particle velocity. Still, in numerical modelling by any reasonable computational code, including certainly the ILSA, the speed equation is always used. (This becomes evident from integration of the SE given in Appendix). The *modified* formulation, in contrast with the conventional formulation, states that the SE is a necessary component of the system of equations and *explicitly* includes the SE into the system. It makes the problem well-posed and defines the propagation of fluid and fracture contours.

*Remark.* For the first time, the special role of the SE in the HF theory was clearly recognized by Kemp [Kemp 1990]. He named it the 'Stefan condition' and wrote (p. 311): "the Stefan condition is always present in moving-boundary problems." Unfortunately, for years, it had not been noticed and employed. It was rediscovered and its fundamental significance was recognized two decades later [Linkov 2011 a,b]. After that, it has been systematically used to obtain analytical and benchmark solutions to HF problems (Linkov 2011, 2012, 2013, 2014 a, b; 2015, 2016 a, b; Mishuris et al. 2012, Linkov and Mishuris 2013, Wrobel and Mishuris 2015]). As concerns with the terminology, note that actually the SE for HF problems has nothing to do with the "*Stefan condition*". The latter describes propagation of a phase transformation front, such as fronts of melting or freezing (see, e.g. https://en.wikipedia.org/wiki/Stefan_problem). As explained in [Linkov 2014 a, 2015], the SE for fluid front is of much more general nature. In contrast with the Stefan conditions, it does not involve energy considerations. It is a consequence of the general Reynolds transport theorem. Its derivation employs merely the axiom of continuity and the concept of particle velocity as the derivative of coordinates of a material point with respect to time. It is true for *any continuous medium* independently on conservation dynamic and constitutive equations. Still, some authors follow Kemp and use the term "Stefan condition" for the fluid SE.

As follows from the Reynolds transport theorem, in the case of non-singular or not too strong singular leak-off, the fluid SE is [Linkov 2014 a, 2015]:

$$v_* = \frac{dx_{*n}}{dt} = \lim_{x \to x_*} v_n \qquad (4)$$

where $v_*$ is the absolute value of the propagation speed at a point $\boldsymbol{x}_*$ of the fluid front, $v_n$ is the normal component of the particle velocity, $\boldsymbol{n}$ is the outward normal to the fluid front at the point $\boldsymbol{x}_*$. In view of (2), the SE (4) reads:

$$v_* = \lim_{x \to x_*} \left(\frac{w^{n+1}}{\mu'}\right)^{1/n} |\nabla p_f|^{1/n-1} \left(-\frac{\partial p_f}{\partial n}\right) \qquad (5)$$

When assuming that at *moving* ($v_* \neq 0$) parts of the front, the fluid velocity has no tangential component ($\boldsymbol{v} = v_n \boldsymbol{n}$), equations (2), (4) and (5) yield the non-trivial consequence that the normal $\boldsymbol{n}$ to the front has the direction opposite to the pressure gradient. Then by (2) and (5), the vector $\boldsymbol{v}_* = v_* \boldsymbol{n}$ of the propagation speed is:

$$\boldsymbol{v}_* = \frac{dx_*}{dt} = -\lim_{x \to x_*} \left(\frac{w^{n+1}}{\mu'}\right)^{1/n} |\nabla p|^{1/n-1} \nabla p \qquad (6)$$

Emphasize that this equation holds merely at points of a *moving* front. It is inapplicable at impenetrable barriers, where $v_n = 0$, and fluid particles, moving along a barrier, have non-zero tangential component. Clearly, there may be intermediate cases, when the fluid velocity at some front points has non-zero both normal and tangential components. Discussion such cases, as well as perturbed motion in a thin boundary layer near a barrier, is beyond the scope of this paper. Equation (6), when applicable, may serve to avoid explicit evaluation of the normal $\boldsymbol{n}$.

The fracture SE (fracture conditions) define the very possibility and the direction of fracture propagation. Commonly, the tensile mode of linear fracture mechanics (see, e.g. [Rice 1968]) is assumed. Then, the fracture conditions are:

$$K_I = K_{IC}, K_{II} = 0, K_{III} = 0 \qquad (7)$$

where $K_I$, $K_{II}$, and $K_{III}$ are, respectively, the normal, shear plane strain and shear anti-plane stress intensity factors (SIFs); $K_{IC}$ is the critical SIF, defined by the strength of a material. The first of (7) defines the very possibility of a fracture growth, two remaining define the direction of the propagation. For a planar fracture, it is assumed that the third of equations (7) is met identically, while the second of (7) conventionally drops out from considerations, because the direction of propagation is assigned along in-plane normal to a front curve.

The SIFs on the left hand side of (7) are factors arising from asymptotic analysis of elastic displacement discontinuities near the fracture contour (e.g. [Rice 1968]). Specifically, $K_I = \sqrt{\pi/32} \, E' \lim_{r \to 0} \frac{w(r,t)}{\sqrt{r}}$. Used in the first of (7), it yields $\lim_{r \to 0} r = \left(\frac{E'}{K'_{IC}}\right)^2 \lim_{r \to 0} w^2$. Then, the temporal derivative of the distance $r$ to the fracture contour gives the absolute value of the contour propagation speed:

$$v_* = \frac{dx_{*n}}{dt} = \lim_{r \to 0} \frac{dr}{dt} = \left(\frac{E'}{K'_{IC}}\right)^2 \lim_{r \to 0} \frac{\partial w^2}{\partial t} \qquad (8)$$

Clearly, equation (8) may be re-written in terms of the dependence of an increment $dr$ of the distance from a fixed near-front point to the front and the increment $dw^2$ of squared opening at this point: $dr = \left(\frac{E'}{K'_{IC}}\right)^2 dw^2$.

Equation (8), being equivalent to the first of (7), makes evident the following statement, which is not distinctly recognized in papers on HF. For a propagating fracture contour, the condition of critical SIF, in essence, presents the speed equation, which defines the main quantity, employed by the theory of propagating interface: the absolute value of the propagation speed.

Underline that the SE for a fluid front (4) and for the fracture contour (8) have quite different originations and they are of quite different generality. As noted, the first of them is a consequence of the Reynold's transport theorem, which is true for any continuous medium, while the second is a consequence of assumptions made in linear fracture mechanics, which are of much less generality. In particular, the simplest conditions (7) are to be modified when a fracture reaches contacts of layers and/or stress barriers (e.g. [He and Hutchinson 1989 a,b]). On whole, in such cases, the fracture conditions are quite uncertain and their reliable specification requires expensive practically unavailable physical experiments. Because of the uncertainty, fracture conditions are assigned by a guess, or by mathematical modeling of local structure and properties, or by playing various scenarios. Then the Neuber-Novozhilov's [Neuber 1943; Novozhilov 1969] approach, which, being a far-reaching extension of the linear fracture mechanics, is preferable for tracing fracture propagation (detailed review and analysis of the fracture criteria of the Neuber-Novozhilov's type may be found in [Dobroskok et al. 2005, Linkov 2010, Sec. 6]). The issue of fracture criteria is also beyond the scope of this paper.

The both SE (5) and (8) are formulated in terms of limits for near front/contour points. It can be shown that near smooth parts of the front and the fracture contour, their right hand sides are defined by the universal 1D asymptotic system of equations, corresponding to 1D problem for a half-infinite fracture under the plane-strain and plane-flow conditions. For the case of a Newtonian fluid and zero leak-off, the asymptotic system, when accounting for the lag, was given by Garagash and Detournay (2000). For the general case of arbitrary power-law fluid and leak-off, when the lag is accounted for, the universal system is derived in the paper [Linkov 2014 b]. Its analysis has confirmed that normally the lag may be neglected. Moreover, for the lag-zone, the distributions of opening and pressure in it may be easily found by using asymptotic matching. It is sufficient to solve a HF problem neglecting the lag (the basic *outer* problem) and to match it with the local solution to the universal 1D asymptotic system (the auxiliary *inner* problem), which accounts for the lag. The matching procedure is described in details in the paper [Linkov 2014 b]. As a result, the lag size and distributions of physical quantities in the lag-zone become known. Estimations, given in [Linkov 2014 b], show that the method of asymptotic matching is acceptable actually in all cases of practical significance. Thus, it is sufficient to focus on the case when the lag is neglected.

For zero lag, the fluid front coincides with the fracture contour. For this case, the significance of asymptotics has been recognized for decades and used for typical regimes of HF propagation. Specifically, for the toughness dominated regime, when the influences of viscosity and leak-off are negligible, the significance of the asymptotic opening has been clearly comprehended in fracture mechanics (see, e.g. [Rice 1968]). It has been also recognized in HF problems since the pioneering works [Spence and Sharp 1985; Desroches et al. 1994; Lenoach 1995]. For particular regimes of fracture propagation, it has been employed in many papers on numerical modeling of HF (e.g. [Lenoach 1995; Adachi and Detournay 2002; Savitski and Detournay 2002; Peirce and Detournay 2008]). Its importance has been specially emphasized in the dedication preceding the paper [Peirce and Detournay 2008]. With the lag neglected, it is reasonable to employ the universal solution, following form the universal system, which depends merely on two dimensionless parameters [Linkov 2014 a; Linkov 2015]. It is applicable to *any regime* of the HF propagation. The asymptotics for limiting [Spence and Sharp 1985; Lenoach 1995], near-limiting [Garagash et al. 2011] and some of transitional [Gordeliy and Peirce 2013; Peirce and Dontsov 2017] regimes are particular cases of the universal solution.

The solution to the universal system with respect to the opening $w$ gives the latter as a function of the distance $r$ and the propagation speed $v_*$ near smooth parts of a front, moving in homogeneous parts of a medium:

$$w = \varphi_w(r, v_*) \qquad (9)$$

The solution (9) presents the *universal asymptotic umbrella* (UAU) [Linkov 2014 a, 2015]. Note that the UAU (9) always identically meets the condition of zero opening on the fracture contour. Using the UAU (9) in (5) is simplified when it has the monomial form:

$$w = A_w(v_*) r^{\alpha(v_*)} \qquad (10)$$

For a general case, the function $A_w(v_*)$ and the exponent $\alpha(v_*)$ are given in [Linkov 2014 a, 2015]. Actually, the exponent $\alpha(v_*)$ changes very slow in a very large range of the propagation speed $v_*$. Specifically, for a Newtonian fluid, $\alpha$ changes from $2/3 = 0.(6)$ to $5/8 = 0.625$, when the propagation regime changes from viscosity dominated to leak-off dominated regime. The difference between the exponents being 1/24, it practically does not influence numerical calculations. However, the dependence of the factor $A_w(v_*)$ on the speed, being much more strong, it is of key influence.

Actually, the UAU (9) presents an implicit form of the SE. The explicit form follows from the UAU (9) by its inversion in $v_*$: $\varphi_v(w(r), r)$. Then the SE becomes

$$v_* = \varphi_v(w(r), r) \qquad (11)$$

with $r = |\boldsymbol{x}_* - \boldsymbol{x}|$ being now the distance from a fixed point $\boldsymbol{x}$ under the UAU to the point $\boldsymbol{x}_*$ on the front along the normal to it. For the monomial dependence (10), $\varphi_v(w(r), r) = A_w^{-1}(w(r)/r^\alpha)$. In practical calculations, when desirable, it is easy to avoid the inversion, because in (9) and (10) $v_* = dr/dt$. Therefore, for an assigned $w$, the UAU (9) becomes a nonlinear ODE in $r$, which is promptly solved in $r$ and consequently in $v_*$. These options are actually employed in the ILSA [Peirce and Detournay 2008]; for a general case, they are explained in Appendix, where efficient solution schemes are suggested.

Substitution (11) into the SE (4) yields

$$v_* = \frac{dx_*}{dt} = \varphi_v(w(r), r) \qquad (12)$$

This completes formulation of the system of equations and boundary conditions. Since the continuity equation (1) and the SE (4) contain the first temporal derivatives of the opening and front location, there is need to assign the initial fracture contour $\psi(\boldsymbol{x}_*, t_0) = 0$ and the initial values of the opening $w_0(\boldsymbol{x})$ within it at the initial time $t_0$. Thus the initial conditions are:

$$\psi(\boldsymbol{x}_*, t_0) = 0, w(t_0, \boldsymbol{x}) = w_0(\boldsymbol{x}) \qquad (13)$$

The fluid front is assumed coinciding with the fracture contour. Finally, the mathematical problem consists of finding the fracture contour $\psi(\boldsymbol{x}_*, t) = 0$ and the opening $w(t, \boldsymbol{x})$ within it from the system (1), (2), (3), (12) for time $t > t_0$ under the initial conditions (13) at $t = t_0$. The boundary condition of zero opening is met identically by the UAU. As mentioned, when having the solution neglecting the lag, the solution for a lag-zone may be promptly found by asymptotic matching.

*Formulation in normalized variables.* Using in (1) and (2) the particle velocity rather than the flux, provides an opportunity to exclude the plane-strain elasticity modulus $E'$ and the viscosity parameter $\mu'$ in a way more convenient than conventional normalizing the time by $(\mu'/E')^{1/n}$. (The latter leads to extremely small units of time). The normalizing is [Linkov 2016 b]:

$$w' = \frac{w}{w_n}, p' = \frac{p}{w_{nE}E'}, \sigma_{n0}' = \frac{\sigma_{n0}}{w_{nE}E'}, q_l' = \frac{q_l}{w_n}, Q' = \frac{Q}{w_n}, K_{IC}' = \frac{K_{IC}}{w_{nE}E'} \qquad (14)$$

with

$$w_n = \left(\frac{\mu'}{E'}\right)^{1/(n+2)} \qquad (15)$$

Then the previous equations hold for the primed variables, while now $E' = 1$ and $\mu' = 1$. Essentially, this normalizing does not change the scales of the time and length. Therefore the time scale still may be chosen as convenient, say with the unit equal to 10 minutes. Moreover, when employing the velocity, rather than the flux, the equations stay the same when all the quantities, proportional to the length scale (coordinates, opening, velocity, propagation speed, leak-off term), are divided by an arbitrary number. This implies that when the fluid injection is simulated as $Q'(t) = Q_0' f_Q(t)\delta(x - x_0)$ by a point source, located at a point $x_0$, the pumping intensity $Q_0'$ may be assumed unit. To this end, in 3D problems, it is sufficient to divide the coordinates, velocity, propagation speed, opening normalized in accordance with (14), and leak-off term by $\sqrt[3]{Q_0'}$, and to divide the critical SIF by $\sqrt[6]{Q_0'}$. In plane problems, the mentioned quantities are divided, respectively, by $\sqrt{Q_o}$ and $\sqrt[4]{Q_o}$. Therefore, without loss of generality, in further discussion we may set $E' = 1$, $\mu' = 1$, $Q_0 = 1$ and consider the instant $t = T_0 = 1$ as corresponding to the initial stage of fracture propagation, say 10 minutes, while $t = T_{Fin} = 100$ corresponds to its final stage (some 1.5 hour). This is also convenient when comparing the results with bench-mark solutions in self-similar variables, because a self-similar solution *a priori* corresponds to $t = 1$. From now on, we assume performed both (i) the normalizing (14), (15) and (ii) subsequent normalizing, which makes the intensity $Q_0$ of the pointed influx unit. The primes in the normalized variables are omitted to simplify notation. *After* the normalizing, the flux $q = wv$ is defined as the product of the normalized opening by the normalized velocity. Below it is used in the discretized continuity equation.

The presented *modified formulation* differs from the conventionally used formulation (e.g. [Adachi et al. 2007, Peirce and Detournay 2008, Peirce 2016, Lecampion et al. 2018]) in three respects. (i) It distinguishes the *particle velocity* as the primary physical quantity; (ii) it explicitly complements the system of equations with the *speed equation*, and (iii) it employs the *universal asymptotic umbrella* to specify the right hand side of the speed equation and to use, when convenient, its inversion in the distance to the front (see Appendix). In numerical calculations, the UAU serves also to assign fluxes on common sides of tip and ribbon elements defined below.

## 3. Spatial discretization. Reduction to system of ODE

*3.1. Spatial discretization. Four collections of elements.* From the problem statement it follows that there are three distinct regions to be distinguished when numerically simulating HF. These are (Fig. 1): (i) central region, where common finite differences are of use; (ii) near-front zone under the UAU, where the UAU is available; and (iii) the front contour to be traced by using methods of the theory of propagating interfaces [Sethian 1999]. In spatial discretization, the three regions involve three distinct groups of mesh elements.

Consider, for certainty, a rectangular mesh covering a region large enough to include the area of front propagation during a time sufficient to pass a number of mesh elements (Fig. 2). The theory suggests distinguishing three major groups of elements, corresponding to the three regions. These are (i) *internal* fracture elements, which have no common sides with elements intersected by the front; (ii) *ribbon* elements, which are within the fracture and have at least one common side with an element intersected by the front; the mesh is assumed to be sufficiently fine; thus the *centers of ribbon elements are under the asymptotic umbrella* (9); (iii) *tip* elements, which are elements intersected by the front. The internal and ribbon elements are completely filled with a fluid; their collection is called *channel* elements. These groups of elements, quite different in computational features, were clearly distinguished and used in the ILSA by Peirce and Detournay [Peirce and Detournay 2008]. We follow the terminology suggested by these authors.

For programming convenience, the three major groups may be complemented with (iv) *external* elements, which are elements outside the fracture. At these elements, the openings, velocities and consequently fluxes are identically zero. As a result, these quantities are zero on their sides, including the sides of those external elements which are neighbours of tip elements.

Therefore, the opening, velocity and flux are zero at a side of a tip element common with an external element.

Denote $\Delta x$, $\Delta y$ the sizes of mesh elements along the $x$ and $y$ axes, respectively. For certainty, the $x$ and $y$ axes are assumed to be horizontal and vertical, respectively, and mesh cells are assigned with numbers $i$ along the $x$-axis from left to right ($i = 1, \dots N_x$), and $j$ along the $y$-axis from bottom to top ($j = 1, \dots N_y$). The opening and pressure are discretized with nodal values $w_{i,j}$, $p_{i,j}$ at centers of cells. Quantities on the left and right sides of a cell $(i, j)$ are supplied with the numbers $(i - 1/2, j)$ and $(i + 1/2, j)$, respectively. Similarly quantities on the bottom and upper sides of a cell $(i, j)$ are assigned with the numbers $(i, j - 1/2)$, $(i, j + 1/2)$.

*3.2. ODE for nodal openings. Identities for mass conservation.* Applying common central differences to (1) and assuming that the pointed source is located at an element $(i_0, j_0)$ yields:

$$\frac{dw_{i,j}}{dt} = \frac{q_{i-1/2,j} - q_{i+1/2,j}}{\Delta x} + \frac{q_{i,j-1/2} - q_{i,j+1/2}}{\Delta y} + q_{li,j} + \frac{Q_0}{\Delta x \Delta y} f(t) \delta(i - i_0, j - j_0) \quad (16)$$

where $q_{i\pm1/2,j}$ and $q_{i,j\pm1/2}$ are $x$ and $y$ components of fluxes on, respectively, the vertical and horizontal sides of mesh cells.

For *external* elements, all the fluxes entering (16) are identically zero; they are also zero at external sides of tip elements. Then multiplication (16) by the area $\Delta S = \Delta x \Delta y$ of a mesh cell and summing over all the cells yields the *identity*:

$$\frac{dV}{dt} \equiv Q_0 \quad (17)$$

where $V = \sum_{i=1, j=1}^{i=N_y, j=N_x} w_{i,j}$ is the total volume of the (incompressible) fluid in a fracture. Thus (17) expresses the global law of mass conservation. To the author's knowledge, this computational identity has not been mentioned in publications on HF (in some papers, their authors even consider the option to complement the system of ODE with the equation of the global mass balance). The identity (17) yields the conclusion, important for further analysis of numerical schemes and numerical results. It is: for the discretization accepted, *the global mass balance is met identically for arbitrary fluxes at sides of mesh cells*. (Recall that the fluxes are set zero at external sides of tip elements).

Clearly, the mass balance (17) implies that for any time interval $(t_1, t_2)$, the increment of the fracture volume $\Delta V$ *identically* equals to the volume of a fluid pumped during this interval

$$\Delta V \equiv \int_{t_1}^{t_2} Q_0 dt \quad (18)$$

The identities (17), (18) may serve to check the accuracy of calculations. In the examples, discussed in Section 8, it has been always met to the accuracy of 6 significant digits, at least. Emphasize again, that the identities are met for whatever fluxes. The latter may be assigned arbitrary; any changes in their prescribing and any errors in their evaluation do not influence the global balance. In numerical calculations, this provides a strong stabilizing effect, which smooths even significant local errors for any scheme of temporal integration.

Return to the ODE (16) and specify the fluxes entering it. In (16), by the definition of the flux as the product of the opening and the particle velocity, we have the fluxes on the right and top sides of an element $(i, j)$:

$$q_{i+1/2,j} = v_{i+1/2,j} w_{i+1/2,j}, \qquad q_{i,j+1/2} = v_{i,j+1/2} w_{i,j+1/2} \quad (19)$$

Herein, $w_{i+1/2,j}$ and $w_{i,j+1/2}$ may be taken as average of nodal values of openings: $w_{i+1/2,j} = (w_{i-1,j} + w_{i,j})/2$, $w_{i,j+1/2} = (w_{i,j-1} + w_{i,j})/2$. The horizontal $v_{i+1/2,j}$ and vertical $v_{i,j+1/2}$ components of the particle velocity in (19) are found by applying central differences to the Poiseuille-type equation (2):

$$v_{i+1/2,j} = -\left\{ w_{i+1/2,j}^{n+1} \left[ \left( \frac{p_{i+1,j}-p_{i,j}}{\Delta x} \right)^2 + \left( \frac{p_{i,j+1}-p_{i,j}}{\Delta y} \right)^2 \right]^{(1-n)/2} \right\}^{1/n} \frac{p_{i+1,j}-p_{i,j}}{\Delta x} \quad (20)$$

$$v_{i,j+1/2} = -\left\{ w_{i,j+1/2}^{n+1} \left[ \left( \frac{p_{i+1,j}-p_{i,j}}{\Delta x} \right)^2 + \left( \frac{p_{i,j+1}-p_{i,j}}{\Delta y} \right)^2 \right]^{(1-n)/2} \right\}^{1/n} \frac{p_{i,j+1}-p_{i,j}}{\Delta y} \quad (21)$$

The fluxes $q_{i-1/2,j}$ and $q_{i,j-1/2}$ on left and bottom sides of the element $(i,j)$ in (16) are found as those for $q_{k+1/2,j}$ $q_{i,l+1/2}$, with $k = i - 1$, $l = j - 1$. For a Newtonian fluid ($n = 1$), the exponent $(1-n)/2$ equals zero, and the non-linear terms in square brackets vanish. Note, that for the case of a non-Newtonian fluid, these terms are defined in (20), (21) not quite precisely, because their entries are evaluated at two different (mutually orthogonal) sides of a cell $(i,j)$. Surely, the accuracy may be improved by using more involved approximation.

As mentioned, actually, the ODE (16), and consequently (19)-(21), refer to nodes in $N_f = N_{int} + N_{rib} + N_{tip}$ elements entirely or partly filled with a fluid; herein, $N_{int}$, $N_{rib}$ and $N_{tip}$ are, respectively, the total numbers of internal, ribbon and tip elements. Note that the equations for external elements may be also included into the system (16), because they are met identically as $0 \equiv 0$. This may serve for programming convenience to keep the number of equations unchanged when a fracture propagates.

In the matrix form, the system of $N_f$ ODE in $N_f$ unknown nodal openings reads:

$$\frac{dw_f}{dt} = \Lambda_f(w_f, p_f)p_f + q_{lf} + \frac{Q_0}{\Delta x \Delta y} f(t)\delta_f \quad (22)$$

where $w_f$, $p_f$ and $q_{lf}$ are, respectively, the vectors of fracture openings, fracture pressures and facture leak-offs; $\Lambda_f(w_f, p_f)$ is a square $N_f \times N_f$ sparse matrix with rows containing at most five non-zero entries, which depend on $w_f$ and, for a non-Newtonian fluid, on $p_f$; $\delta_f$ is the source vector with the only non-zero component equal to 1 for the element $(i_0, j_0)$ containing the pointed source.

The nodal values $p_{i,j}$ of the net-pressure, entering (20), (21), are defined from the discretized hypersingular equation (3). The openings in it are assumed piece-wise constant when performing integration. Then integration over rectangular elements with piece-wise constant openings, associated with centers (nodes) of elements yields:

$$p_{i,j} = \sum_{k,m \in C_f} G_{ijkm} w_{k,m} + \Delta\sigma_{i,j} \quad (23)$$

where $C_f$ is the collection of numbers assigned to $N_f$ fracture elements. Again, if wanting to have a system of ODE with unchanged number of equations, the summation in (23) may be taken over the entire mesh, because the openings in external elements are zero.

The influence coefficients $G_{ijkm}$ in (23) are (e.g. [Linkov and Mogilevskaya 1986; Linkov, Zoubkov and Heib 1997; Peirce and Detournay 2008]):

$$G_{ijkm} = \frac{1}{8\pi} \left[ \frac{\sqrt{(x_i - x_{k+1/2})^2 + (y_i - y_{m+1/2})^2}}{(x_i - x_{k+1/2})(y_i - y_{m+1/2})} + \frac{\sqrt{(x_i - x_{k-1/2})^2 + (y_i - y_{m-1/2})^2}}{(x_i - x_{k-1/2})(y_i - y_{m-1/2})} - \right.$$

$$-\frac{\sqrt{(x_i-x_{k-1/2})^2+(y_i-y_{m+1/2})^2}}{(x_i-x_{k-1/2})(y_i-y_{m+1/2})}-\frac{\sqrt{(x_i-x_{k+1/2})^2+(y_i-y_{m-1/2})^2}}{(x_i-x_{k+1/2})(y_i-y_{m-1/2})}\Bigg] \qquad (24)$$

Equations (23), (24) provide the pressure on the entire mesh including external elements, as well. Condensing equations (23) to $N_f$ fracture elements ($i,j \in C_f$), they become:

$$p_f = G_f w_f + \Delta\sigma_{of} \qquad (25)$$

where $G_f$ is the square $N_f \times N_f$ fully populated matrix of influence coefficients, $\Delta\sigma_{of}$ is the vector of stress contrasts at fracture elements. Again, if wanted, the matrix $G_f$ may be extended on the entire mesh: the pressure at nodes of external elements are not used in the ODE (16), because the velocities, openings and consequently fluxes at their sides are set zero.

It should be noted that the pressure in *tip* elements is found extremely inaccurate what drastically influences the fluxes on their boundaries with *ribbon* elements. This issue is left for special discussion in Sections 4, 6 and 8.

Successive substitutions (24) into (23), the latter into (20), (21), the result into (19) and the latter into (16) give, after using the vector forms (22) and (25), the system of $N_f$ ODE in $N_f$ openings at nodes on the fracture:

$$\frac{dw_f}{dt} = \Lambda_f\big(w_f, p_f(w_f)\big)G_f w_f + \Lambda_f\big(w_f, p_f(w_f)\big)\Delta\sigma_{oc} + q_{lf} + \frac{Q_0}{\Delta x \Delta y}\delta_f \qquad (26)$$

Note that the right hand sides of the system (26), besides the openings, depend also on the front position. The latter changes in time and it defines (i) the fluxes from ribbon to tip elements and (ii) collections of tip, ribbon, internal and external elements. Therefore, the system (26) is to be complemented with speed equations, which govern the front propagation.

*3.3. Speed equations.* These equations are obtained by writing the SE for centres of ribbon elements which are assumed to be under the UAU (10). The inversion in the module of the propagation speed (11) yields the SE associated with the centres of ribbon elements. For $N_R$ ribbon elements, this gives $N_R$ ODE for the modules of propagation speeds $v_{*j}$ associated with each of them:

$$\frac{dr_j}{dt} = v_{*j} = \varphi_v(w_{Rj}, r_j) \qquad (27)$$

where $w_{Rj}$ is the opening at the center of the $j$-th ribbon element ($j = 1,\ldots,N_R$), $r_j$ is the distance from the center to the fracture front.

Underline that the SE (27) are separate in the sense that each of them includes values merely for a single ribbon element. Actually, for each ribbon element, we have a single ODE with the right hand side often given analytically. This drastically simplifies their integration. For an explicit scheme, it may be performed either by direct substitution the known at the beginning of a time step values $w_{Rj}$ and $r_j$ into the right hand side of (27), or by using commonly available inversion of the UAU in the distance. For an implicit scheme, the integration is promptly and highly efficient performed by using either the function $\varphi_v(w_{Rj}, r_j)$, presenting the inversion of the UAU (9) in the speed, or alternatively, by using the inversion of the UAU in the distance $r_j$, or by using the function $\varphi_w(r, v_*)$ itself. Details are given in Appendix.

The current distances $r_j$, obtained by integration (27), define *circles* centered at nodes of ribbon elements. Their *envelope* presents the current fracture front and consequently it gives the normal at each its point. Therefore, for a *fixed collections* of ribbon and tip elements, the system of ODE (26), (27) completely defines the evolution of the nodal openings and of the front.

Building the envelope may be replaced by or complemented with *tracing* movement of *selected points on the front (marker particles)* at each of the tip elements. These points can be defined through a chosen parametrization of the fracture contour at the beginning of a time step. For such a point, the movement direction, coinciding with the normal, is known, while *the speed magnitude is assigned via the speeds magnitudes defined by* (27) *for neighboring ribbon element(s)*. When a tip element has more than one neighbor, the mean value of speeds may be used. Then the movement of fixed (marker) front points may be traced in the course of integration (explicit or implicit) the SE (27). This provides additional data on the front changes, which opens additional options for its reconstruction on each (for implicit schemes) or selected (for explicit schemes) time steps. Specifically, instead of (or in addition to) building the envelope to the circles, defined by integration of (27), the simplest marker/string method may be efficiently employed when a fracture contour expands. Details of the two approaches to tracing the front (via the envelope to circles or/and via marker particles) are discussed in Sec. 5.

*3.4. Joined system of ODE.* Thus the problem is reduced to integration the joined system of ODE (26), (27) and tracing the front under known nodal openings and front position at the beginning of a time interval, within which the collections of tip, ribbon, internal and external elements are the same. The collections are to be checked and updated from time to time. The updating of the collections becomes needed only when some of current tip elements becomes a ribbon element. Thus it is reduced to checking from time to time, if a front still intersects a current tip element. The examination, being a simple purely geometrical problem, it is briefly discussed afterwards.

The principle issues to be discussed are (i) *assigning the fluxes* at common sides of tip and ribbon elements, (ii) tracing the front propagation, defined by the SE (27), and, perhaps the most significant, (iii) making a choice between *time stepping* schemes. Consider these three issues.

## 4. Assigning fluxes on common sides of tip and ribbon elements

It is of essence, that the pressure in tip elements is found extremely inaccurate: the errors are of order of tens and even hundreds percent. This unavoidably leads to wrong values of the pressure gradient, which defines the velocities (20), (21) and consequently fluxes (19) on internal sides of tip elements. The errors further aggravate when the second derivatives of the pressure, present in the flux divergence, are evaluated in the continuity equations (16) for tip elements and also for their neighbor ribbon elements. Thus special attention is to be paid to proper assigning the fluxes on common sides of tip and ribbon elements.

There are two options: either (i) to use extremely inaccurate values of the pressure in tip elements, when the pressure is defined by applying the hypersingular operator to openings, or (ii) to avoid their using by means of the UAU. Consider these options.

*4.1. Finding fluxes into tip elements by using the pressure: statistical method.* This option has been avoided when solving truly 3D problems by well-established codes like the ILSA (e.g. [Peirce and Detournay 2008, Peirce 2016]). Meanwhile, recently Stepanov [Stepanov 2018] has disclosed that this path is not forbidden when employing multiple small time steps for explicit integration of the continuity equations (26). For the KGD benchmark problem, the study of *average* values on a time interval, during which the front propagates a mesh element, has confirmed that errors in the tip pressure are very large (up to thousands percent). However the particle velocity at the boundary with neighboring ribbon element has notably less errors (within hundreds percent). What is of special significance, the velocity changes monotonically and passes through the exact value near the middle of the interval. The errors of the opening increment in the ribbon element, obtained by integration of (26), are much less; they are on the level of at most first percent depending on the mesh size [Stepanov 2018]. Finally, the accurate evaluation of openings, entering the right hand sides of the speed equations (27), results in quite accurate tracing the front propagation, as well.

Similarly good results have been obtained when temporal integration of the ODE was performed by using implicit integration. The latter permits large time steps, while it requires iterations within a time step. Then favorable averaging occurs over successive iterations. Since for both explicit and implicit integration, the accurate values appear as average over a number of time steps or iterations, the approach is named the statistical method [Stepanov 2018]. Underline, that for a planar fracture, this approach avoids using the normal to the front. The direction of the normal is implicitly accounted for by the direction of evaluated pressure gradient. The latter is collinear to the normal in the area under the UAU, within which the centers of ribbon and tip elements are assumed to be.

The unexpectedly favorable results, provided by the statistical method, may be explained by the mentioned smoothing influence of the global mass balance (18). Recall that the balance is met identically on any time interval and for arbitrary great errors in instant values of the flux.

*4.2. Direct assigning fluxes via the universal asymptotic umbrella.* The asymptotic umbrella, which covers the centers of ribbon elements, allows one to avoid using unreliable tip pressure. The UAU serves both to find the particle velocity at points of a common side of tip and ribbon elements and to evaluate the average opening along this side. Their product is the needed flux.

Consider a ribbon element $(i, j)$ and a tip element, with which it has a common vertical or horizontal side. The part of the fracture front, which intersects the tip element, has known normal $\boldsymbol{n}$ with the components $n_x$ and $n_y$ at its points. Assume, for certainty, that the simplest piece-wise linear approximation is used for the front, so that the normal has the same direction along a front segment. When, as assumed in (6), the propagation speed is collinear to the normal ($\boldsymbol{v}_* = v_*\boldsymbol{n}$), the propagation speed and particle velocities under the asymptotic umbrella have the same direction. Since the common side of a tip and a ribbon element is under the UAU, the normal $\boldsymbol{n}$ and the speed magnitude $v_*$ are known. Specifically, the normal is that to the front at its part intersecting the tip element. The speed magnitude is that defined by the SE (27) for the ribbon element. In its turn, the average opening on the common side may be found via the UAU (9) either by means of integration (9) along the side, or simply as the value, defined by (9) at its middle point. In particular, when using the last option, the flux on a common vertical side is

$$q_{i\pm1/2,j} = n_x v_* \varphi_w(v_*, r_{i\pm1/2}) \tag{28}$$

where the upper (lower) sign is taken when $n_x > 0$ $(n_x < 0)$. Similarly, when the common side is horizontal, the flux on it is

$$q_{i,j\pm1/2} = n_y v_* \varphi_w(v_*, r_{i,j\pm1/2}) \tag{29}$$

where the upper (lower) sign is taken when $n_y > 0$ $(n_y < 0)$.

*4.3. "Hidden" assigning fluxes via the asymptotic umbrella by using apparent pressures.* Direct employing the UAU, used in (28), (29), is the simplest option to assign fluxes without using the tip pressure evaluated via the hypersingular operator. The authors of the ILSA [Peirce and Detournay 2008] employed another path. In subsection 4.4.4 of their paper, they used the asymptotic umbrella to find the volume, and consequently the mean opening $w_t$ of a tip element, at the end of a time step. For a tip element, its mean opening is evaluated via the UAU (9) as:

$$w_t = \frac{1}{\Delta x \Delta y} \int_{S_t} \varphi_w(v_*, r) dS \tag{30}$$

where $S_t$ is the part of the tip element behind the front; this part is filled with the fluid. The speed magnitude $v_*$ is assumed found through the speeds defined by the SE for the neighbor ribbon elements. When approximating the front by straight segments, the integral on the right hand side of (30) is reduced to the contour integral over the contour $L_t$ of the part $S_t$. Then (30) becomes:

$$w_t = \frac{1}{\Delta x \Delta y} \int_{L_t} \Phi_w(v_*, r(\eta)) d\eta \tag{31}$$

where $\Phi_w(v_*, r)$ is an antiderivative to $\varphi_w(v_*, r)$ in the distance $r$; $r$ and $\eta$ are the local Cartesian coordinates with the $r$-direction opposite to the normal $\boldsymbol{n}$ $(n_x, n_y)$ to the front and with the $\eta$-axis along the front segment. In the case of a monomial asymptotics (10), the antiderivative is $\Phi_w(v_*, r) = A_w(v_*) r^{\alpha+1}/(\alpha + 1)$ and the contour integral in (31) is evaluated analytically [Peirce and Detournay 2008]. In 1D problems, the result is obvious: $w_t = A_w(v_*)(x_* - x_{i\pm1/2})^{\alpha+1}/(\alpha + 1)$, where the upper (lower) sign corresponds to the right (left) tip of a straight fracture; $i$ is the number of the ribbon element neighboring the tip element considered.

With known $w_t$, the temporal derivative $dw_t/dt$ for the tip element considered is also known. In particular, for the conventional first order approximation, we have $dw_t/dt = (w_t - w_{t0})/\Delta t$, where $w_{t0}$ is the known opening at the beginning of a time step of duration $\Delta t$. Then substitution this approximation into the left hand side of the conservation equation (26) for the tip element gives actually the needed total influx into this element from neighbouring ribbon elements. Clearly the global mass balance (17) is met identically. The authors slightly complicate calculations along this path by introducing an 'apparent' tip pressure, which provides the needed influx through the finite differences (20), (21), involving these pressures. Thus the 'apparent' tip pressures become additional unknowns defined by conservation identities for each of tip elements.

In essence, this approach is close to direct prescribing the fluxes in accordance with (28), (29). In fact, in *one*-dimensional problems, when the direction of fracture propagation is predefined, the approaches are equivalent. Our comparative experiments, performed for 1D benchmark KGD problem, solved (similar to [Peirce and Detournay 2008]), by using the implicit Euler scheme, confirmed this. However, for a *two*-dimensional planar fracture, in cases, when a tip element has *two* sides common with *two* ribbon elements, the approach by [Peirce and Detournay 2008], in contrast with (28), (29), does not distinguish *two* influxes into the tip element; it provides their sum through the *single* apparent pressure. The *apparent* pressure is defined by the global mass balance; it replaces the unreliable tip pressure, defined by the elasticity equation (23).

The direct assigning the fluxes into tip elements in the line of (28), (29) does not involve additional unknowns and it does not require rearranging the system (26) to find them. Thus this option may be preferable, especially, when integrating the ODE (26) explicitly. In all the cases, the global mass balance (18) is met identically.

## 5. Tracing the front propagation: reconstruction of contour and updating the collections of elements

The ODE (26), (27) are written for fixed collections of the tip, ribbon, internal and external elements. In particular, equations (27) are formulated for $N_R$ ribbon elements. Consider firstly the front propagation within a time interval $t_{\Delta x}$ during which these collections do not change. The duration of this interval is of order $\Delta T = \Delta z/v_*$, where $v_*$ is a current propagation speed and $\Delta z$ is the minimal mesh size. Within this interval, tracing the front consists of *reconstruction of front contour* on each of selected time or iteration steps under fixed collections. When the time approaches the end of the interval, there arises the need in *updating the mentioned collections*. Thus, there are two distinct parts of front tracing: (i) *reconstruction* of the contour under fixed collections of elements, and (ii) *checking and updating* these collections. The second part, being actually trivial, we shall focus on the front reconstruction. Updating the collections is briefly discussed at the end of the section.

For the front reconstruction, we have merely $N_R$ separated SE (27) associated with $N_R$ ribbon elements. Integration of each of them, discussed in Appendix, defines at the end of each time step (i) the distance $r_j$ from the center of $j$-th ribbon elements to the front, and (ii) the propagation speed $v_{*j}$ ($j = 1,..., N_R$). These data may serve, for two options to reconstruct the front contour at the end of a time or iteration step. These are, respectively, (i) defining the new contour as the envelope to $N_R$ circles of radius $r_j$, centered at centers of ribbon elements, and (ii) finding the new contour by usting the final positions of the traced front points. Surely, these options may be combined. Consider their simplest realizations.

*5.1. Reconstruction the contour via the envelope to circles. Connection with Eikonal equation.* Building an envelope to a finite number of circles is purely geometrical problem, which may be solved to various degree of smoothness of the envelope curve. The simplest approximation is piece-wise linear; it approximates the envelop by straight segments tangential to two successive circles (Fig. 3). Their centers 1 and 2, numbered counter clockwise, are located at the centers of successive ribbon elements. The radii of the circles 1 and 2 are $r_1$ and $r_2$, respectively. The absolute value of their difference is $\Delta r = |r_1 - r_2|$. The vector $\boldsymbol{d}$, connecting the centers, starts at the point $\boldsymbol{x_1}$ with coordinates $x_1, y_1$ and ends at the point $\boldsymbol{x_2}$ with coordinates $x_2, y_2$; thus the length of $\boldsymbol{d}$ is $d = \sqrt{(x_2 - x_1)^2 + (y_2 - y_1)^2}$. The tangent, being external to the fracture, it should be $d > \Delta r$; hence, $c = d/\Delta r > 1$. The normal $\boldsymbol{n}$ to the envelope is directed to the right to the direction of $\boldsymbol{\tau}$. The angles of vectors $\boldsymbol{d}$, $\boldsymbol{\tau}$ and $\boldsymbol{n}$ with the $x$-direction are, respectively, $\gamma_d$, $\gamma_\tau$ and $\alpha = \gamma_\tau - \pi/2$. Clearly, it is sufficient to find the angle $\alpha$ of the normal $\boldsymbol{n}$ with the $x$-direction to have the start $\boldsymbol{\tau_1} = \boldsymbol{x_1} + \boldsymbol{n}r_1$ and end $\boldsymbol{\tau_2} = \boldsymbol{x_2} + \boldsymbol{n}r_2$ tangent points.

Note firstly that from the geometrical picture, there follow three inequalities, connecting angles $\alpha$ and $\gamma_d$: 1) $\gamma_d - \alpha \leq \pi/2$ if $r_1 \geq r_2$; 2) $\gamma_d - \alpha \geq \pi/2$ if $r_2 \geq r_1$; and 3) $\gamma_d - \pi \leq \alpha \leq \gamma_d$. Then it can be shown that:

$$\tan \alpha = \frac{\mp\sqrt{c^2-1}+\tan\gamma_d}{1\pm\tan\gamma_d\sqrt{c^2-1}}$$

$$\cos \alpha = sign(\cos\alpha)\left|\frac{\cos\gamma_d}{c}(1\pm\tan\gamma_d\sqrt{c^2-1})\right| \qquad (32)$$

$$\sin \alpha = sign(\sin\alpha)\left|\frac{\cos\gamma_d}{c}(\mp\sqrt{c^2-1}+\tan\gamma_d)\right|$$

Herein, the upper (lower) signs are taken, when $r_1 \geq r_2$ ($r_2 \geq r_1$); $c = d/\Delta r$, $\tan\gamma_d = (y_2 - y_1)/(x_2 - x_1)$, $\cos\gamma_d = (x_2 - x_1)/d$. The signs of $\cos\alpha$ and $\sin\alpha$ are uniquely defined by the sign of $\tan\alpha$ and the inequalities 3).

With known $\boldsymbol{\tau_1}$ (or $\boldsymbol{\tau_2}$) and $\boldsymbol{n}$, the distance (if needed) from the tangent line to an arbitrary point is defined by elementary geometry. In particular, if the circles correspond to the ribbon elements, shown in Fig. 3b ($x_2 - x_1 = -\Delta x$, $y_2 - y_1 = \Delta y$, $r_2 \geq r_1$), then $\tan \alpha = \frac{\sqrt{c^2-1} - \Delta y/\Delta x}{1 + (\Delta y/\Delta x)\sqrt{c^2-1}}$, $\quad \cos \alpha = \frac{\Delta x/d}{c}(1 + \sqrt{c^2-1}\,\Delta y/\Delta x)$, $\quad \sin \alpha = \frac{\Delta x/d}{c}(\sqrt{c^2-1} - \Delta y/\Delta x)$.

The distance from the envelope line to the center of the tip element $\boldsymbol{x}_{tip}$ with the coordinates $x_1$ and $y_2$ is: $r_3 = r_2 - \Delta x \cos \alpha$. These results coincide with those derived for the scheme of Fig. 3b in the paper [Peirce and Detournay 2008; p. 2871, 2872]. The authors of [Peirce and Detournay 2008] obtained them by solving the Eikonal equation in the line of the theory of propagating interfaces [Sethian 1999]. They gave another geometrical interpretation to the solution, which does not involve the envelope to the circles centered at ribbon elements. Thus the derivation above shows that using finite differences on a rectangular mesh to solve the Eikonal equation corresponds to piece-wise linear approximation of the envelope. Clearly, the concept of the envelope to circles suggests a wide variety of alternative options for its approximation.

*5.2. Reconstruction the contour via marker particles.* As mentioned at the end of subsection 3.3, the solution to the SE (27) provides also possibility for easy and fast tracing the movement of selected points (marker particles) on the front. Assume for certainty that merely a single front point is fixed and traced in each of tip elements belonging to a given collection of tip elements. Then the number of marker particles equals to the current total number $N_{tip}$ of tip elements. Numerate the points counter-clockwise. The *i*-th point $\boldsymbol{x}_i$ has the global coordinates $x_i, y_i$ ($i = 1,\dots, N_{tip}$).

The simplest reconstruction of the contour may be performed by line segments as: $\boldsymbol{x}_i = \lambda \boldsymbol{x}_i + (1 - \lambda)\boldsymbol{x}_{i+1}$ ($0 \le \lambda \le 1$) with $\boldsymbol{x}_{N_{tip}+1} = \boldsymbol{x}_1$. The components $n_{ix}, n_{iy}$ of the unit normal $\boldsymbol{n}_i$, corresponding to the *i*-th marker particle, are reconstructed via the central differences as [Sethian 1999]: $n_{ix} = -(y_{i+1} - y_{i-1})/\Delta s)$, $\quad n_{iy} = (x_{i+1} - x_{i-1})/\Delta s)$, where $\Delta s = \sqrt{(x_{i+1} - x_{i-1})^2 + (y_{i+1} - y_{i-1})^2}$.

This approach actually corresponds to the standard *marker/string method* of front reconstruction. The numerical tests discussed below show that it provides results to the accuracy of the solution to Eikonal equation when the front expands as a convex curve. In general, the two approaches, both being not time expensive, may complement each other.

*5.3. Updating the collections of elements.* Clearly there is no need to perform updating at each time step, especially, when using explicit integration, for which a time step is very small. Actually, it is sufficient to perform updating after a number of time steps. A rough estimation of the time interval between successive updating is given above as $\Delta T = \Delta z/v_*$, where $v_*$ is a current propagation speed and $\Delta z$ is the minimal mesh size. The exact instant to start updating is found from the following considerations.

The collections change only when the front advances and intersects a new *external* element, which becomes a *tip* element. Therefore, it is sufficient to check when the reconstructed front intersects a new external element. With known analytical description of the contour on a time step, it is a simple geometrical problem. When using marker particles in each tip cell, it is even simpler to examine if a particle is still in the same cell. Therefore, there is no sense to dwell on details of the updating.

## 6. Considerations for the choice between explicit and implicit integration of ODE

A system of ODE under prescribed initial conditions may be solved by using a wide variety of well-established methods (e.g. [Epperson 2002]). As known, the methods are of two major classes: (i) *explicit*, which provide the values at the end of a time step merely through values already found on previous step(s); and (ii) *implicit*, which involve in calculation the values to be found as an output of a time step. For certainty, we shall assume Euler forward and backward schemes as representative for, respectively, explicit and implicit integration.

A proper choice between these quite different options strongly depends on a particular problem. Therefore, it is reasonable to discuss the properties of the HF system of ODE (26), (27). They are different for the internal part of the fracture and for its near-front zones.

*6.1. Properties of the ODE for the central part of fracture.* At the central part of the fracture, in contrast with the near-front zones, the solution (openings at nodal points) changes relatively slow. Then using average (over the considered part) value of the displacement when calculating the matrix $\Lambda_f(w_f, p_f(w_f))$, the system (26) becomes of the form

$$\frac{dw}{dt} = \Lambda_f(w_{av})G_f w + F(t) \tag{33}$$

where $\Lambda_f(w_{av})G_f$ is a matrix with frozen coefficients, which strongly depend on the minimal mesh size, say $\Delta x$. As shown in the paper [Peirce 2006] by the analysis of a 1D problem for a Newtonian fluid, the condition number of the matrix $\Lambda_f(w_{av})G_f$ in the Euclidean metric has the order $1/\Delta x^3$ (the product of: (i) $1/\Delta x$ in hypersingular elasticity operator in (3), (ii) $1/\Delta x$ in the pressure gradient in the Poiseuille equation (2), and (iii) $1/\Delta x$ generated by the divergence in the continuity equation (1)). This implies that the system (26) for *the central part is stiff*, what has drastic computational consequences for both explicit and implicit schemes of time stepping.

*6.1.1. Consequences of high stiffness for explicit schemes.* For *explicit* schemes, the Courant-Friedrich-Levi (CFL) stability condition requires the time step $\Delta t_{exp}$ not exceeding the maximal admissible value $\Delta t_{max} = K_{CFL}\Delta x^3$, where factor $K_{CFL}$ depends on a particular HF problem. Thus the stability condition for the time step $\Delta t_{exp}$ of an explicit scheme is:

$$\Delta t_{exp} \leq K_{CFL}\Delta x^3 \tag{34}$$

From the estimations, given in [Peirce 2006] for eigenvalues of the matrix, it may be deduced that the factor $K_{CFL}$ in (34) is of order $0.5 t_n/w_{av}{}^3$, where $t_n = (\mu'/E')^{1/n}$ is the only parameter with the time dimension present in the non-normalized HF equations (1)-(3). The exact value of $K_{CFL}$ for a particular problem may be found from numerical experiments. Specifically, as appears from calculations, discussed in Sec. 8, for the bench-mark KGD problem, solved in variables normalized in accordance with (14), the factor $K_{CFL}$ at the instant $t = T_0 = 1$ is around 0,46. Similar order is also established in numerical experiments for the normalized axisymmetric problem. Then the CFL condition (34) obtains the form:

$$\frac{\Delta t_{exp}}{T_{ref}} \leq 0.46\Delta\varsigma^3$$

where $T_{ref}$ is the reference time, at which the estimation is made, $\Delta\varsigma = x/x_*$ is the mesh size normalized by the fracture half-length $x_*$. This estimation is quite instructive in practical calculations. We use it for preliminary conclusions on the time expense of explicit schemes.

We use this estimation for preliminary conclusions on the time expense of explicit schemes.

From the estimations it follows that for a fine mesh with say 100 cells along a fracture half-length (radius), the time step should be of order $10^{-7}$ of the initial (normalized) time $T_0 = 1$, and, consequently, more than $10^7$ steps are required to reach the normalized time $t = 2$. As clear from (33), each step involves multiplication of the fully populated matrix $G_f$ by the opening vector $w$ with $N_f$ components, what, for fine meshes, leads to extreme time expense if the multiplication is performed without acceleration, or/and parallel computation. Meanwhile, for a rough mesh with merely five elements along the initial half-length (radius), the time expense is four orders less. Such time cost may be acceptable even without accelerations for calculations on conventional laptops.

It is of significance to compare computational complexities of explicit and implicit integrations. For explicit schemes, an estimation immediately follows from the CFL condition (34). It implies that the minimal number of time steps, needed to trace the distance of the mesh size $\Delta x$ with the speed $v_*$, is $1/(v_* K_{CFL} \Delta x^2)$. The corresponding number of arithmetic multiplications $C_{exp}$ is $N_f{}^2$ greater:

$$C_{exp} = N_f^2/(v_* K_{CFL} \Delta x^2) \tag{35}$$

Below in subsection 6.3 it is compared with the complexity of implicit integration.

*6.1.2. Consequences of high stiffness for implicit schemes.* The great advantage of implicit integration is the possibility to use a large time step. Still, for *implicit* schemes, in their turn, high stiffness of the system (33) leads to another unfavorable consequence. Although the time step may be much greater than for an explicit scheme, the price for this benefit is the need to solve a strongly non-linear algebraic system. Its solution is found iteratively by a variant of fixed-point iterations, say, by the Newton's method. Within an iteration in non-linearity, the coefficients of the matrix are frozen. The unfavorable feature of the matrix is that because of high condition number, proportional to $1/\Delta x^3$, the matrix is *ill-conditioned*. Then solving the linear system may become unstable if not using a specially designed preconditioner (e.g. [Epperson 2002]). Construction of a proper preconditioner is of essence for both exact and iterative methods of solving the system. For a large system, arising in truly 3D problems, the solution is found iteratively. Then properly constructed preconditioner is needed to find an accurate solution with non-excessive number of iterations.

Thus, in addition to the usual for implicit schemes need to solve a strongly non-linear system involving re-building its matrix, there arises the *need in building of a proper preconditioner* for a current frozen matrix. This problem has been tackled and successfully solved for a Newtonian fluid in the papers [Peirce and Siebrits 2005, Peirce 2006]; the preconditioner is employed in the ILSA [Detournay and Peirce 2008, Peirce 2016]. Still the problem requites further work as regards to non-Newtonian fluids.

Each of iterations still involves the multiplication of a fully populated matrix by a vector of iterated openings. Thus, the total count $C_{imp}$ of multiplications, needed to trace the front propagation to the distance of the minimal mesh size, say $\Delta x$, is

$$C_{imp} = N_f^2 n_{sts} n_{ext} n_{int} \tag{36}$$

where $n_{sts}$ is the number of time steps of implicit integration to trace the distance $\Delta x$, $n_{ext}$ is the number of fixed point iterations, in which the non-linear operator is approximated by a matrix with frozen coefficients, $n_{int}$ is the number of iterations needed to solve an algebraic system with frozen coefficients by an iterative method with properly chosen preconditioner. As appears from the paper [Peirce 2005], $n_{int} = 7$ when using the preconditioner suggested in this paper. The value $n_{ext} = 6$ is a favorable estimation for the number of fixed-point iterations. The number of needed steps $n_{sts}$ may be 4 or even 2. When taking the conservative value $n_{sts} = 4$, it appears that for a properly constructed preconditioner, the product $n_{sts} n_{ext} n_{int}$ does not exceed 200.

*6.2. Properties of the ODE for the near-front zones of fracture.* On whole, at areas where the opening does not change fast, the system of starting partial differential equations (1)-(3) is parabolic. This is especially evident for the PKN model [Perkins and Kern 1961, Nordgren 1972], when the opening is proportional to the pressure. Then substitution (2) into (1) yields the right hand side with the second spatial derivative of the opening, while the left hand side contains the first temporal derivative.

However, because of non-linearity of the Poiseuille-type equation (2), containing the degree $w^{n+1}$ of the opening, the *opening changes very fast* near the boundary of the region. This yields that the equation parabolic at the central part, becomes hyperbolic near the end zone. Zeldovich and Kompaneets [Zeldovich and Kompaneets 1950] were the first to distinguish such a feature of equations describing thermal processes with non-linear thermal conductivity. These authors disclosed the wave-like propagation of the heat front when the coefficient of thermal conductivity grows proportionally to a degree of the temperature. They obtained a self-similar solution in separated spatial and temporal variables without *a priori* suggestion on the moving boundary: its propagation appears as a mathematical result. It can be seen that, when the conductivity is proportional to the *second* degree of the temperature, the equations for the thermal problem are exactly the same as those for the Perkins-Kern-Nordgren PKN model [Perkins and Kern 1961, Nordgren 1972] when neglecting the leak-off. Therefore, the results by Zeldovich and Kompaneets refer to the PKN problem, as well. As to the author's knowledge, the similar peculiarity has not been distinctly exposed for HF problems, although tracing the fracture front in time was the essence of numerous publications.

The next two subsections contain general considerations on this issue and the mathematical proof that the wave-like propagation also follows from the benchmark solutions, obtained in self-similar variables, to the KGD and axisymmetric HF problems. The analysis provides also estimations of the size of the zone, which moves nearly as a simple wave; the estimations are used for conclusions on numerical modeling of HF.

*6.2.1. Wave-like propagation of the near-front zone.* For a general HF problem, the features of the propagation can be revealed by considering the temporal derivative $\frac{\partial w}{\partial t}\big|_x$ present in the continuity equation (1) under fixed global coordinates. Evaluate it in a near-contour zone by using the local coordinate $r$, counted from the contour in the direction opposite to the outward normal. Assuming the global $x$–axis to be along the normal, the partial derivative $(\partial w/\partial t)_{x=const}$ is connected with the partial derivative $(\partial w/\partial t)|_r$, evaluated under fixed $r$, as

$$\frac{\partial w}{\partial t}\Big|_x = \frac{\partial w}{\partial t}\Big|_r + \frac{dr}{dt}\frac{\partial w}{\partial r} \qquad (37)$$

Since the opening is identically zero at the contour, the first term on the right hand side of (37) goes to zero when $r \to 0$. In contrast, as follows from the UAU (10), the second term goes to infinity when the derivative $dr/dt$ is non-zero. This derivative is the instant speed $v_*$ of a point on the contour. Therefore, if $v_* = dr/dt \neq 0$, then in the global coordinates, the opening at a near-front zone meets the wave equation:

$$\frac{\partial w}{\partial t} = -v_* \frac{\partial w}{\partial x} \qquad (38)$$

where it is assumed that the global $x$-axis is taken in the direction of the normal to the front at its point considered. Emphasize that this result is not trivial as may appear when deriving the asymptotic umbrella starting from the *assumption* that the near-front zone propagates steadily with a constant velocity (e.g. [Descroches et al. 1994]). In fact, obtaining asymptotic umbrella does not require this assumption. It is especially evident from self-similar solutions. For instance, the authors of the pioneering work [Spence and Sharp 1985] solved the HF equations in separated variables, which provided the self-similar solution to the KGD problem as the product of a function only of time to a function merely on the normalized coordinate. An assumption on wave-like propagation is not needed and naturally it was not used. They obtained the asymptotic umbrella from purely asymptotic analysis for the viscosity dominated regime without the mentioned assumption. There is also no need in this suggestion when obtaining the general form of the UAU. As shown in [Linkov 2014 a, 2015], the fact that the asymptotic equations for a near-front zone formally correspond to its steady propagation is the deduction (not an assumption) of the asymptotic analysis. The derivation above is free of this assumption, as well.

The wave-like propagation is not *a priori* suggestion. As clear from the derivation, the wave equation (38) is the consequence of (i) *zero opening at the fracture contour, and (ii) singular behavior of its spatial derivative evident from the UAU.*

*6.2.2. Consequence of wave-like propagation for acceptable mesh sizes.* It is of essence to study to which accuracy the linear wave solution $w(x,t) = w(x - v_*t)$ of (38) is met at the centre of a ribbon element. Recalling that the latter should be under the UAU, this will provide an estimation of an acceptable mesh size.

Self-similar solutions to the KGD and axisymmetric problems may serve to this purpose. Emphasize that they present purely mathematical results free of *ad hoc* suggestions on asymptotics and change of opening in time. In the both cases, for a power-law fluid, the exact solution is found in separated temporal and spatial variables. In the normalized coordinates, the solution is [Adachi and Detournay 2002; Savitski and Detournay 2002; Linkov 2016 c]:

$$x_* = \xi_{*n}t^{\gamma_x}, \; w = \xi_{*n}W(\varsigma)t^{\gamma_w}, W(\varsigma) = C_n(1-\varsigma)^\alpha[1 + a_1(1-\varsigma) + \cdots] \tag{39}$$

where $\varsigma = x/x_*$ is the spatial variable; $\xi_{*n}, \gamma_x, \gamma_w, \alpha, a_1$ are constants depending merely on the fluid behavior index $n$; they are defined in the papers cited.

Denote $t_1$ and $t_2$ the instances, when the centres of the current ribbon elements 1 (at $t_1$) and 2 (at $t_2$) are at the same distance $r = \Delta x$ from the front. Denote $w_1$ the opening of the element 1 at $t_1$; $w_2$ the opening of the element 2 at $t_2$. Then for the ratio $w_2/w_1$, the solution (39) yields:

$$\frac{t_2}{t_1} = \left(\frac{x_{*2}}{x_{*1}}\right)^{1/\gamma_x}, \quad \frac{w_2}{w_1} = \left(\frac{x_{*1}}{x_{*2}}\right)^\alpha \left[1 + O\left(\left(\frac{\Delta x}{x_{*1}}\right)^2\right)\right]\left(\frac{t_2}{t_1}\right)^{\gamma_w}$$

Substitution the first of these equations into the second and accounting for $x_{*2} = x_{*1} + \Delta x$, gives:

$$\frac{w_2}{w_1} = \left(\frac{x_{*1}}{x_{*2}}\right)^\beta \left[1 + O\left(\left(\frac{\Delta x}{x_{*1}}\right)^2\right)\right] = 1 - \beta\frac{\Delta x}{x_{*1}} + O\left(\left(\frac{\Delta x}{x_{*1}}\right)^2\right) \tag{40}$$

where $\beta = \alpha - \gamma_w/\gamma_x$. For a straight fracture, the definitions of $\alpha, \gamma_w, \gamma_x$ [Adachi and Detournay, 2002] give $\beta = 1/(n+1) - 2/(n+2)$; for a penny shaped fracture, the definitions [Linkov, 2016 c] yield $\beta = 0.5(2-n)/(n+1) - 2/(n+2)$. The exponent $\beta$ is maximal for a Newtonian fluid ($n = 1$) being $\beta = 1/6$ for a straight fracture, and $\beta = 7/12$ for a penny-shaped fracture. For a perfectly plastic fluid ($n = 0$), $\beta = 0$ in the both cases.

Equation (40) implies that to the accuracy of the term $\beta\Delta x/x_{*1}$, the distribution of the opening near the front propagates the mesh size $r = \Delta x$ as a wave with unchanged form ($w_2 = w_1$). Therefore, to the accuracy at least 3.3% for a straight fracture and 11.7% for a penny-shaped fracture, this occurs even when the number of channel elements on a fracture half-length (radius) $x_*$ is merely five ($\Delta x/x_* = 1/5$). Direct calculation of the ratio $w_2/w_1$ by using the exact values of $w$, shows that the estimation is rather conservative. Actually, even for so rough mesh ($\Delta x = x_*/5$), the opening distribution near the front stays unchanged to the accuracy of less than 3%.

This implies that, when accounting for the asymptotics, quite a rough spatial mesh may be used for modeling a *near front* zone of HF, where the opening changes *fast*. Clearly, a rough mesh is applicable, as well, at the *central* part of a fracture, where the opening varies *slowly*. Therefore, *quite rough spatial meshes may serve for numerical modeling of HF*. This theoretical conclusion, important for simulation of HF, agrees with the results of numerical calculations performed by using quite rough meshes in examples, given in the papers [Peirce and Detournay 2008; Peirce 2016] for a Newtonian fluid. Our calculations for both Newtonian and non-Newtonian fluids, confirm this conclusion, as well. Recall also that, as established in the previous section, for rough meshes, the time expense of an explicit scheme is acceptable even without accelerations.

*6.2.3. Consequence of wave-like propagation for the time step and mesh size.* The disclosed fact that the asymptotic umbrella propagates a mesh size nearly as a simple wave, is quite favourable for using explicit schemes. For the wave equation (38), the CFL stability condition is

$$\Delta t_{exp} \leq \Delta x / v_* \tag{41}$$

In contrast with the CFL condition (34) for the central part, containing cube of the mesh size $\Delta x$, the condition (41) is much less restrictive containing merely the first degree of $\Delta x$. Specifically, for the solution (39) to the self-similar problems, $v_* = dx_*/dt = \gamma_x \xi_{*n} t^{\gamma_x}$, where, as follows from numerical data of [Adachi and Detournay 2002; Linkov 2016 c] for thinning fluids ($0 \leq n \leq 1$), the factor $\gamma_x \xi_{*n}$ is within the interval from 0.24 to 0.40. Then for the instant $t = T_0 = 1$, the time step may be as large as $2.5 \Delta x / x_*$. Even for a fine mesh with 100 elements along the half-length ($\Delta x / x_* = 0.01$), the acceptable time step is 0.025. It is five orders greater than the step required by the condition (34) for the central part. Clearly it is not restrictive.

We conclude that the condition (34) for the central part, being much more restrictive, the time step for uniform meshes is controlled by this condition. Yet, *equation (41) justifies using fine meshes for the near-front zone*. This becomes especially favorable for simulation of propagation through a stress barrier. Then the speed $v_*$, entering the denominator on the right hand side of (41), drastically decreases, and the stability condition becomes even more favorable.

This conclusion raises the question regarding to the concern expressed in the review [Lecampion et al. 2018, p. 72] that "an implicit (for fluid-solid coupling) / explicit (for fracture front increment) scheme will perform poorly in the presence of heterogeneities and/or in-situ stress jumps if the time-step is too large". The question is: if such schemes will indeed "perform poorly"?

From the conclusion above it follows that when a fracture front has already penetrated a barrier and propagates in it with a small speed, a fine mesh near the front may be used even for a large time step in frames of *explicit* integration. For the central part, nothing prevents *implicit* integration with this time step. Hence, in this case, a hybrid scheme looks acceptable.

Quite different and much more involved matter is the simulation of the penetration process for a strong barrier. Consider for simplicity one dimensional KGD or axisymmetric problem, when the entire fracture front simultaneously reaches high plane or circular stress barrier. At the instant, when the barrier is reached, the fracture propagation speed and, consequently, the particle velocities near the front abruptly turn to zero. Then the velocity component orthogonal to the channel sides dominates over the velocity component along the sides; the local movement in the channel becomes unsteady and inertial terms, employed in the paper [Cao, Hussain, Schrefler 2018], may become of significance. Therefore, the key postulates, assumed when deriving a Poiseuille-type equation, are violated. Obviously, no solution based on a Poiseuille-type equation can quantitatively describe the fluid motion when a fracture reaches a very high (impenetrable in the limit) barrier.

When not accounting for these complicating factors, one may only hope on more/less satisfactory qualitative description of the physical picture. The simulations of reaching and penetration into a strong stress-barrier, performed Gladkov and Linkov [Gladkov and Linkov 2017, 2018] for the extended KGD problem, have disclosed the reverse movement with high oscillations about a small (practically zero) value of the particle velocity. On the main part of the fracture, the oscillations rapidly disappear; near the barrier the velocities themselves very quickly become close to zero. The fracture blows up under arrested tip until the average pressure increases to a level sufficient to overcome the barrier. A very fine local mesh was used to model the near-barrier zone and the barrier itself. For a fine mesh on the entire fracture, the explicit Runge-Kutta integration was as stable as the Petzold–Gear BDF implicit method, when the time step for the explicit scheme was taken according the CFL condition for the central part. These results indicate, while not prove, that hybrid schemes with fine local meshes near a barrier may serve for modeling the penetration.

However, for extremely high (impenetrable) barriers, it was established that explicit, as well as implicit back Euler and Petzold-Geer BDF methods, are unstable. In this case, only Brayton-Gustavson-Hachtel BDF method [Brayton et al. 1972] provided stable physically consistent results. A confident answer to the question of proper choice of integration schemes, when a fracture reaches a rigid wall, is left to future theoretical and numerical investigations.

*6.2.4. Consequence of wave-like propagation for schemes of spatial discretization of the near-front zone.* The wave-type propagation, described by (38), influences also *schemes* of spatial discretization for a near-front zone. It implies that the spatial mesh should met the *causality principle*: the information for the solution should propagate in the direction of the front movement. Thus *upwind* spatial schemes are preferable to provide physically consistent and computationally stable results.

This is especially important for *explicit* time stepping, if the time step is chosen to meet the highly restrictive condition (34) on entire fracture. In this case, the time step is quite small, and tracing the propagation of a near front zone on a typical distance of the mesh size requires a large number of steps. Then spatial *backward* finite differences are superior over central differences for approximation spatial derivatives near the front (e.g. [Sethian 1999]). In practice, for *explicit* schemes, as concerns with the ODE (26), (27), this implies that *it is safer not to use*: (i) the openings of those tip elements, whose centers are ahead of the front; only activated elements (those already passed by the front) are taken into account; and (ii) the asymptotic equations (28), (29) for the fluxes on common sides of tip and ribbon elements.

The first recommendation is an obvious consequence of the causality principle. The second follows from the theoretical considerations, which show that using (28), (29) in explicit stepping leads to computational instability on multiple very small time steps. The instability is caused by exponential amplification of a small perturbation $\delta w$. It arises because equations (28), (29) combined with the speed equation (27) actually involve dependences of the *rewind* form $dw/dt \approx Cw/\Delta T$ with *positive* $C$ of order 1 and with the time $\Delta T = \Delta x/v_*$ for propagating the distance $\Delta x$ of the mesh size.

This specific difficulty of HF problems when using *explicit* temporal integration is easily overcome by proper modification of the spatial difference scheme. For non-activated tip elements, the fluxes on their common sides with ribbon elements are set zero. For activated tip elements, the fluxes on their common sides with ribbon elements are defined by equations (19) with the only change: the velocities $v_{i+1/2,j}$ and $v_{i,j+1/2}$ are set equal to the respective components $v_{*x} = v_* n_x$ and $v_{*y} = v_* n_y$, of the propagation speed $\boldsymbol{v}_*$. As usual, the fluxes on the sides of tip elements with external elements are set zero. Speed equations (27) stay the same.

These minor improvements of the spatial discretization exclude numerical instability. Numerous numerical experiments for 1D problems, performed by the author with various mesh sizes and with various schemes of *explicit* integration, have shown that neglecting any of the items (i) and (ii) may lead to deterioration of the solution within the time, for which the front propagates the distance of three elements, at most. In contrast, when properly accounting for these requirements, the results are accurate and stable in a wide (two and more orders) interval of time even under strong perturbation of initial conditions. The calculations discussed below for explicit integration are performed in accordance with the recommendations (i) and (ii).

*6.3. Comparison of time expenses for explicit and implicit stepping.* Before making overall conclusions on a preferable time stepping, it is reasonable to compare the time expenses $t_{exp}$ and $t_{imp}$, needed to trace the fracture advance to the distance of a mesh size. They are proportional to the complexities $C_{exp}$ and $C_{imp}$, defined by (35) and (36), respectively. Thus the ratio of time expenses is:

$$\frac{t_{imp}}{t_{exp}} = \frac{C_{imp}}{C_{exp}} = (K_{CFL} \Delta x^2 v_*)(n_{sts} n_{ext} n_{int}) \qquad (42)$$

Notably, *the estimation* (42) *does not depend on the number of unknowns*. Therefore, detailed comparison of time expenses may be made by focusing on the simplest one-dimensional benchmark problem. As has been mentioned, from the estimation given in [Peirce 2006] for the KGD model, it follows that the factor $K_{CFL}$ is of order $0.5 t_n/w_{av}{}^3$. In the normalized variables used in (39), this yields $K_{CFL} = 0.5/(\xi_{*n} W_{av})^3$ at a typical time $T = 1$. In these variables, $\Delta x^2 v_* = \gamma_x \xi_{*n}^3 \Delta \varsigma^3$, where $\Delta \varsigma = \Delta x/x_*$. Then in the normalized variables, at a typical time $T = 1$, we have $K_{CFL} \Delta x^2 v_* = 0.5 \gamma_x \Delta \varsigma^2/W_{av}{}^3$.

It can be shown that by using the global mass balance, that $W_{av} = 1/(2\xi_{*n}^2)$ for the KGD problem, and $W_{av} = 1/(2\pi\xi_{*n}^3)$ for the axisymmetric problem with $\xi_{*n}$ given, respectively, in the papers [Adachi and Detournay 2002] and [Linkov 2016 c] for fluids with the behaviour index $n$. Consider, for certainty, a Newtonian fluid ($n = 1$, $\gamma_x = 2/3$) and the KGD problem ($\xi_{*n} = 0.615$). In this case, $W_{av} = 1.3$. Our direct study of the stability for the KGD model, solved by the explicit Euler scheme, gives a bit less value $W_{av} = 1.04$ of the averaged opening, corresponding to the stability threshold. With the latter result, the estimation of the product $K_{CFL} \Delta x^2 v_*$ becomes

$$K_{CFL} \Delta x^2 v_* = \Delta \varsigma^2/3 \qquad (43)$$

The number of matrix-to-vector multiplications for the explicit scheme is inverse to this quantity; it is $3/\Delta \varsigma^2$. For a rough mesh with merely 5 elements along a half-length, the number is 75 only. However, it grows rapidly for finer meshes, and becomes $3 \cdot 10^4$, when the mesh size becomes 0.01 of the fracture half-length.

Recall now the estimations for the implicit (backward Euler) scheme, given in the normalized variables for the same benchmark problem: $n_{sts} = 4$, $n_{ext} = 6$, $n_{int} = 7$. The number of matrix-to-vector multiplications, needed to trace the fracture advance to the distance of a mesh size, is given by their product entering (42): $n_{sts} n_{ext} n_{int} = 168$. Notably, this number is two-fold greater, than that obtained above for explicit integration with a rough mesh (75 such multiplications).

With the estimation (43) of $K_{CFL}\Delta x^2 v_*$ for explicit integration and the estimation $n_{sts}n_{ext}n_{int} = 168$ for the implicit scheme, the ratio (42) becomes

$$\frac{t_{imp}}{t_{exp}} = \frac{c_{imp}}{c_{exp}} = 56\Delta\varsigma^2 \qquad (44)$$

From (44) it follows that the time expense for an explicit scheme is less (greater), than that for an implicit scheme, when the mesh size in the central part of a fracture is greater (less) than 0.13 of its half-length. Therefore, if the number of elements over a fracture half-length does not exceed 7, the explicit scheme requires even less time, than the implicit scheme. When having 10-25 elements along the half-length, the time expense of an explicit scheme is still acceptable being merely 2-11-fold greater than that of an implicit scheme. However, the time expense of an explicit scheme (without accelerations of matrix-to-vector multiplication) becomes two orders greater than that of an implicit scheme, when $\Delta\varsigma$ is less than 0.013 (about 80 elements along the half-length).

Above it has been established that a rough mesh with merely five elements along the half-length is applicable for tracing the fracture propagation in a homogeneous medium. Now it appears that for such a rough mesh, the time expense of explicit integration is comparable with implicit integration.

Thus, we have come to the key conclusion. Contrary to the common opinion, *the time step of explicit integration is not prohibitively small* in problems of HF. With this result on the time step, it is of essence to compare other features of explicit and implicit integration.

*6.4. Properties of the ODE for the near-source zone.* In the continuity equation (1) and its discretized form (16), the pumped influx $Q_0$ per unit time is prescribed by the Dirac's delta-function. This implies that in a 3D case, the flow near the source is axisymmetric, and the radial flux $q_r$ goes to infinity as $q_r = Q_0/(2\pi r)$ with deceasing distance $r$ from the source point $(x_0, y_0)$. Hence, the particle velocities in (20), (21) and the fluxes in (19) and (16) go to infinity at sides of the source cell $(i_0, j_0)$ when mesh sizes decrease. In practical calculations, the last (positive) term on the r. h. s. of (16) rapidly (as $Q_0/(\Delta x\Delta y)$) grows what leads to proportional growth of the magnitude of two first (negative) terms. This means that actually the r. h. s. of (16) is evaluated as uncertainty of the type $(-\infty) + (+\infty)$. Then for a fine mesh the accuracy unavoidably deteriorates even when using arithmetic of double precision. Simple estimations and direct calculations show that the deterioration may occur for quite a rough mesh of some 5-10 elements along a fracture half-size.

Clearly, this is an artificial effect caused by the representation of the source by the delta-function. On the other hand, it is unreasonable to complicate the problem by accounting for details of the local geometry of a borehole. Its influence rapidly disappears when the size of a hydraulic fracture becomes greater that 5 diameters of a borehole [Zubkov et.al. 2006].

The simplest remedy to overcome the difficulty may consist of employing the mentioned asymptotics of the flux by prescribing its averaged components at sides of the source cell. Following this line, we have for the asymptotic fluxes $q_x$ and $q_y$ on the vertical and horizontal side, respectively,

$$q_x = \frac{Q_0\Delta x}{4\pi(y^2 + 0.25\Delta x^2)}, \quad q_y = \frac{Q_0\Delta y}{4\pi(x^2 + 0.25\Delta y^2)}.$$

For fluxes $q_{xav}$ and $q_{yav}$, averaged over the cell sides $\Delta y$ and $\Delta x$, respectively, integration yields

$$q_{xav} = \frac{Q_0}{\pi\Delta y}\operatorname{atan}\frac{\Delta y}{\Delta x}, \quad q_{yav} = \frac{Q_0}{\pi\Delta x}\operatorname{atan}\frac{\Delta x}{\Delta y}$$

The total influx, defined by these fluxes, identically meets the mass balance: $2q_{xav}\Delta y + 2q_{xav}\Delta x \equiv Q_0$. For a square mesh ($\Delta x = \Delta y$), the average fluxes are $q_{xav} = q_{xav} = 0.25 Q_0/\Delta x$. The average fluxes $q_{xav}$ and $q_{yav}$ may be used in (16) for cells adjusting the source cell ($x_0, y_0$). Then the ODE for the cell ($x_0, y_0$), containing the singular point, becomes of no need, and it drops out from the system (16). The corresponding opening is found by extrapolation of openings at the neighboring cells. This opening is used when finding the pressure via the elasticity equation (25).

## 7. Explicit versus implicit integration. Conclusions on the choice between schemes

With the conclusion of the previous section that, despite strongly restrictive CFL, the time expense of explicit schemes is not prohibitively small, we may compare the explicit and implicit integration in other computational respects. Recall the advantages and drawbacks of the two choices.

The *key advantage of an implicit scheme* is the possibility to use a large time step. A mesh element may be passed in two-four time steps only. However, this beneficial feature is reached for the price of *disadvantages*. There arises the need to solve a strongly non-linear system. Its solving includes two distinct difficulties. The first consists of the need to develop an appropriate fixed-point operator to linearize the non-linear system, say, by the Newton method. The iterations involve successive rebuilding a fully populated matrix, performed in the external cycle of calculations. Secondly, on each of these iterations, it is necessary to solve a linear system of algebraic equations for the stiff system of the starting ODE. For a large system, the solution is found iteratively in the internal cycle of iterations. To overcome the stiffness, they require building a proper preconditioner, which provides convergence with *non-excessive* number of iterations.

To the date, these complications have been successfully overcome for a Newtonian fluid by the authors of the ILSA [Peirce and Detournay 2008; Peirce 2016]. Still, there is need in new tools to solve problems for non-Newtonian fluids to compete with and/or complement implicit integration. As appears from the results of the previous section, explicit integration suggests such options.

The *major advantages of explicit schemes*, when applied to the HF system (26), (27), are:

(i) they do not require inversion of matrices; merely matrix-to-vector products are used;

(ii) the main fully populated matrix, used in matrix-to-vector multiplications, is fixed and evaluated in advance;

(iii) they are of immediate applicability to non-Newtonian fluids with arbitrary rheology;

(iv) high stability to large changes of initial conditions, given the time step is small enough;

(v) easy drastic acceleration of matrix-to-vector multiplications by using fast multipole methods (for arbitrary meshes) and fast Fourier transform (for uniform meshes).

(iv) easy implementation of parallel computations for the multiplications.

The *key disadvantage of explicit integration*, as noted in many papers and emphasized above, consists of the need to meet the highly restrictive CFL condition (34) for a time step. However, as has been revealed by the analysis, this restriction is far from being fatal.

The general conclusions concerning with the choice are:

(i) explicit schemes are superior over implicit schemes, when the time expenses are comparable; this is the case for rough meshes on the central part of a fracture;

(ii) explicit integration may complement/replace implicit integration, when the time expense of an explicit scheme, although being notably greater than that of an implicit scheme, is still acceptable.

## 8. Numerical results for bench-mark problems

The conclusions of the previous section are verified numerically by solving two benchmark problems, for which accurate (with at least four correct digits) solutions are available. These are the KGD problem and the HF axisymmetric problem for a penny-shaped fracture. Their solutions (39) have been already employed for obtaining general estimations. The KGD problem contains all the germs of generality except for the need to find the direction of the fracture propagation: the fracture propagates along the $x$-axis; the direction is predefined. The specific issue of finding the direction of front propagation concerns actually with the front reconstruction considered in Sec. 5. It will be commented by using the exact solution for a penny-shaped fracture.

*8.1. Numerical results for the KGD problem.* Since according to (42) the *ratio* of time expenses for explicit and implicit integrations does not depend on the number of unknowns, the KGD problem provides easy opportunity to compare explicit and implicit schemes numerically. Of essence is that the maximal number $N_c$ of unknown openings does not exceed first hundreds. When comparing implicit and implicit methods, the fluid was assumed Newtonian, because to the date, the implicit scheme has been applied merely to this case. Yet for explicit schemes, in view of their easy applicability to fluids with arbitrary rheology, solutions for non-Newtonian fluids were studied, as well.

Without loss of generality, the calculations were performed in the normalized variables (14), (39) under constant pumping rate starting from the initial time $T_0 = 1$ to the final time $T_{\text{Fin}}$. In the most part of calculations the latter was set $T_{\text{Fin}} = 100$. In some cases it was taken much less (when comparing the time expense per mesh size), or an order greater (when studying the accuracy and stability). The mesh size $\Delta\varsigma$ was assigned as a fracture $1/N$ of the initial half-length $x_{*0}$ with $x_{*0} = 0.6118$ for a Newtonian fluid. The number $N$ of elements on the initial half-length was varied from 4 to 40.

The explicit scheme does not involve solving non-linear algebraic system. Therefore, there is no need to comment on its implementation, except for the choice of the time step in accordance with the CFL condition (34). On the other hand, for an implicit scheme, it is of essence to comment on its implementation as concerns with fixed-point iterations used to solve the system (26), (27) on a large time step. Consider these issues.

*8.1.1. Choice of time step for explicit method.* For explicit time stepping, we employed the methods of forward Euler, Runge-Kutta and explicit Adams. Direct numerical tests confirmed that, in accordance with the theoretical CFL condition (34), all of them required a small time step $\Delta t$ of order $O(\Delta x^3)$ to be stable. A typical dependence of the maximal $\Delta t$, which still provides stability, on the mesh size $\Delta\varsigma$ is plotted in log-log coordinates in Fig. 4 for the forward Euler method. The dependence may be approximated as $\Delta t = 0.46\Delta\varsigma^3$. Hence, in the considered 1D problem, the factor $K_{CFL}$ in (34) equals 0.46.

For a fixed spatial size $\Delta x$, there was no difference in the accuracy of results when using various explicit methods. The time expense was also on the same level. This implies that the forward Euler method, being the simplest, may serve as an acceptable choice. The results for explicit time stepping over a distance of the mesh size were obtained by this method with the step $\Delta t = 0.46\Delta\varsigma^3$. However, in calculations simulating the propagation over a large time interval ($T_{Fin} = 100$), for safety, the time step was taken five-fold less than the critical value. This served to account for the fact that growing number of elements of unchanged size $\Delta x$ corresponds to a finer mesh on a final (at $T_{Fin} = 100$) fracture half-length $x_*$, because the ratio $\Delta\varsigma = \Delta x / x_*$ decreases.

*8.1.2. Comparison of time expenses.* Calculations by the explicit and implicit Euler schemes have shown that, in agreement with the estimations of the previous section, for a rough mesh with 5 elements on the initial half-length ($\Delta x / x_{*0} = 0.1223$), the time expenses are quite close, while the explicit scheme works some faster. For instance, tracing the front propagation to the distance of the mesh size $\Delta x$ required the time 0.0167 s, when using the explicit method. For the implicit method, it was, respectively, 0.034 s and 0.069 s, when using merely two and four time steps to trace the same propagation. Thus, for the rough mesh ($\Delta\varsigma = 1/5$), explicit integration requires 2-3-fold less time than implicit integration. For two-fold finer mesh $\Delta x / x_{*0} = 0.06118$ ($\Delta\varsigma = 1/10$), the time expense of the explicit integration grew some 4-fold. Then the time expenses of explicit and implicit become practically the same.

*8.1.3. Comparison of accuracy of temporal integration under fixed mesh size.* The accuracy of *explicit* stepping is actually defined merely by the spatial mesh size, because in view of quite small time step, temporal integration is performed very accurately. The results of integration by using schemes of high order of accuracy, such as Runge-Kutta method of fifth order, evidently confirm this. In contrast, the accuracy of an *implicit* scheme with large time step depends on the time step. Therefore, in practice, the results, obtained by an explicit method, may serve to evaluate the accuracy of implicit time stepping for large steps of various duration.

The comparison is performed when tracing the front propagation to the distance of mesh size $\Delta x$ starting from the initial time $T_0 = 1$. For a coarse mesh with 5 elements on the initial fracture half-length ($\Delta x / x_{*0} = 0.1223$), the explicit *forward* Euler method required about 384 steps. As has been mentioned, it provided high accuracy of temporal integration. Consequently, it may serve as a bench-mark for a given spatial mesh to estimate the accuracy of implicit schemes for the same mesh. Implicit integration by *backward* Euler method with merely two and four time steps to trace the same propagation, had the errors of -2.03% and -1.7%, respectively. Taking into account that for the coarse mesh the error, caused by the spatial discretization, is 0.56%, the error of the implicit method with large steps is some 4-fold greater.

For two-fold finer mesh ($\Delta x / x_{*0} = 0.06118$), the explicit method, as expected, required about 4-fold greater number of steps (about 1500). Implicit integration for this mesh had the errors some three-fold smaller than those for the rough mesh. They were - 0.8% and - 0.5% when using two and four time steps, respectively. For this mesh, the error caused by spatial discretization, is 0.065% only. Therefore, the error of implicit integration with large steps, while quite small, is yet an order greater than that of the explicit method.

The conclusions on the accuracy are as follows. Any explicit scheme gives the results coinciding with the bench-mark solution when the latter is obtained on the same mesh. The errors of an implicit scheme with large time steps, although being greater, are acceptable.

*8.1.4. Comparison of sensitivity to strong perturbation of initial conditions.* The results above are obtained when using the initial conditions provided by the exact bench-mark solution at the instant $T_0 = 1$. Surely, this is a very favorable starting guess for solving the problem with iterations on a time step. As shown in [Linkov 2016 a, b], even under very strong perturbations of the initial conditions at $T_0 = 1$, the solution with growing time tends to that given by the self-similar solution for a pointed source. Since in practice initial conditions are quite uncertain, it is of essence to check if the two approaches compared are able to overcome the uncertainty and to provide correct results at final time, say, $T_{Fin} = 100$.

The study of sensitivity to perturbations was performed for the initial conditions of the form:

$$x_*(1) = x_{*0}, w(x, 1) = \varepsilon_w w_0(x) \qquad (45)$$

where $x_{*0}$ is the half-length provided by the self-similar solution, $w_0(x)$ is the corresponding distribution of the opening, $\varepsilon_w$ is the perturbation parameter. For $\varepsilon_w = 1$, the distribution is that of the self-similar solution; it is quite favourable for time stepping. For $\varepsilon_w = 0.1$ and $\varepsilon_w = 0.01$, we have this distribution strongly perturbed. It has appeared that both explicit and implicit schemes are quite insensitive to strong perturbations of the initial conditions. Still, the explicit Euler scheme is more stable to the perturbations: it allows an order greater perturbations.

*8.1.5. Applicability of explicit schemes to non-Newtonian fluids.* The major disadvantage of explicit schemes, which is a quite small time step, turns into its advantage as regards to dealing with non-linearity. For a small time step, a scheme becomes equally applicable to Newtonian and non-Newtonian fluids. Calculations for the benchmark problem considered have confirmed that actually there is no principal difference when applying explicit Euler scheme to Newtonian ($n = 1$) and thinning fluids $(0 < n < 1)$. The differences concern with merely factors and powers in the Poiseuille-type (2) and speed (5) equations.

*8.2. Numerical results for the axisymmetric problem.* An application of the explicit method to a planar fracture, propagating in 3D elastic medium, is tested by considering the axisymmetric propagation of a penny-shaped fracture under constant pumping rate. The solutions to the accuracy of four correct significant digits, at least, are given in the papers [Savitski and Detournay, 2002] for a Newtonian fluid, and [Linkov 2016 c] for any power-law fluid.

Calculations were performed for a square mesh with the side $\Delta x$. The latter was chosen to have $N$ elements on the initial fracture radius $x_{*0} = r_{*0}$ ($\Delta x = x_{*0}/N$). For $N = 5$, the mesh is quite rough; it contains some 100 square channel elements on the initial fracture. For $N = 10$, the mesh is two-fold finer; it contains some 400 channel elements at the initial time. The variables, being normalized in accordance with (14), (39), the initial time is $T_0 = 1$ and the intensity of the source is $Q_0 = 1$. Then for a Newtonian fluid, $x_{*0} = r_{*0} = 0.6978$ [Savitski and Detournay 2002; Linkov 2016 c]. The bench mark solution, used for the comparison, was obtained in self-similar variables by a special subroutine in accordance with the algorithm described in [Linkov 2016 c].

*Time expense for explicit scheme.* Taking into account the estimation (35), it may be expected that the time expense will be $N$-fold greater for the penny-shaped fracture than for a straight fracture with the same number of elements along an initial fracture half-length. It would grow similar to 1D case as $N^2$ when the mesh size $\Delta x$ decreases proportionally to $1/N$ ($\Delta x = x_{*0}/N$). Detailed calculations, performed by A. Stepanov in close collaboration with the author, have shown that the conclusions of subsection 8.1 for the KGD problem regarding to the temporal and spatial steps, accuracy, insensitivity to strong perturbations of initial conditions and applicability to non-Newtonian fluids refer to the penny-shaped HF, as well.

In particular, the explicit code is of immediate use for non-Newtonian fluids. The graphs in Figures 5 and 6 present the profiles of the opening for, respectively, a Newtonian ($n = 1$) and typical thinning ($n = 0.6$) fluids obtained by using the explicit Euler scheme (circular markers). For comparison, the benchmark solutions are shown by solid lines. In may be seen that agreement of the numerical results with exact openings is quite good, while the accuracy is a bit better for a Newtonian fluid.

We conclude that the explicit method provides results to the accuracy sufficient for engineering applications with acceptable time expense. This refers to calculations on a conventional laptop even without using acceleration of matrix-to-vector multiplications and without parallel calculations for this multiply repeated operation. The calculations for non-Newtonian fluids are performed similar to those for a Newtonian fluid. They do not require special adaptation of the algorithm.

*Reconstruction of the front.* The problem for a penny-shaped fracture, when solved as a planar problem, may serve to compare various methods to reconstruct the fracture front considered in Section 6. The comparison of the marker/string method with the fast marching method of the Eikonal type, employing piece-wise linear envelope to the circles, has been performed by A. Stepanov in collaboration with the author. The conclusion for the convex contour is: the simplest string/marker method provides the accuracy similar to that of the fast marching method.

*Accuracy of spatial integration and finite differences.* It is quite illuminating to study the accuracy, provided by the conventional *spatial discretization* with using the piece-wise constant openings in the elasticity equation and central differences when calculating the pressure gradient, particle velocities and divergence of the flux. To this end the exact bench-mark solution to the axisymmetric problem may be substituted into the right-hand sides of the discretized elasticity (24), Poiseuille-type (20), (21) and continuity (16) equations. The results are to be compared with the exact bench-mark values of the left-hand sides of these equations. We performed such a study for a Newtonian fluid for square meshes of two sizes. Detailed numerical results are obtained and analysed for meshes with $N_m = 20$ and $N_m = 40$ cells along the fracture horizontal diameter. In the first case the total number of cells covering the penny-shaped fracture was 357; in the second 1337.

It is established that the discretized elasticity equation (24) provides the pressure to the accuracy on the level of 3% for $N_m = 20$, and of 1% for $N_m = 40$, except for nodes of ribbon elements. At ribbon nodes, the singular behavior of the solution near the front appears in growing errors, especially, for the finer mesh ($N_m = 40$), for which the ribbon nodes are closer to the front. The error grew to 12% and 30%, respectively, for $N_m = 20$ and $N_m = 40$.

The openings on the *sides* of cells evaluated as the arithmetic average of exact openings at two neighbour nodes, are more accurate than the pressure. They have 3-4 correct significant digits, except for common sides of ribbon and tip elements. At these sides, strongly influenced by the asymptotics, the errors again grew faster for the fine mesh. They were some 20% and 30% for the numbers $N_m = 20$ and $N_m = 40$, respectively.

Pressure gradient, found by using central differences for sides of cells, has much greater errors than the pressure. The error reaches 60% at the sides of the source cell and at the common sides of tip and ribbon elements. On sides, which are out of strong influence of singularities, the error of gradient is on the level of, respectively, 5% and 10% for meshes with $N_m = 20$ and $N_m = 40$.

The particle velocities, defined by central differences (20), (21), and fluxes defined by equations (19), have the accuracy similar to that of pressure gradient.

The greatest errors, naturally, occur when evaluating the divergence of the fluxes defined by central differences on the right-hand side of the discretized continuity equation (16). For both $N_m = 20$ and $N_m = 40$, the values of the divergence near the source and near the front are absolutely wrong. They differ orders from the bench-mark values near the source, and 2-3-fold under the asymptotic umbrella, including the centres of ribbon elements. At remaining nodes, the errors are 10-50% for $N_m = 20$, and 1-10% for $N_m = 40$.

The conclusion is: on a time step, the right-hand sides of the ODE (26) are evaluated very inaccurately. Using fine meshes even aggravates the errors at cells near the source point and near the front, which are under strong influence of singular asymptotics.

Nevertheless, these large errors on a particular time step do not manifest themselves in the final solution obtained after temporal integration. As clear from     the   results   of   temporal integration, discussed above, when the front propagates the distance of a mesh size, the errors of openings, pressure and front positions are on the level of percent even for a quite rough mesh with merely 10 cells along a diameter. For finer meshes with $N_m = 20$ and $N_m = 40$, the errors become less than 1%. This is due to random character of errors on time or/and iteration steps. During the time interval of propagation a mesh size, the poorly evaluated quantities oscillate about the exact values, so that after their integration over time, the results become quite accurate. Finally, the main quantities, which are of interest for HF, are calculated to the accuracy acceptable for practical applications.

Therefore, temporal integration smooths large errors of flux divergence on individual time steps. The Stepanov's statistical method {Stepanov 2018} vividly illustrates the smoothing effect of temporal integration. Surely, the smoothing effect is essentially due to the outlined fact that the system of ODE (16) and its matrix forms (22), (26) identically meet the mass conservation law (17) on each time step and, consequently, as stated by (18), on any time interval.

## 9. Summary

The conclusions of the research are as follows.

(i) A proper formulation of HF problems should include speed equations (SE) as a necessary component of the system. The SE for this class of problems arise either as the *consequence* of Reynolds transport theorem for points at a fluid front, or in the form of a fracture condition, *postulated* for points at a fracture contour. *Complementing the system with the SE makes a HF problem complete and well-posed*. The formulation employs global spatial coordinates and hypersingular operator connecting the opening with the fluid pressure. Employing the global coordinates and the SE clearly reveals singular behavior of the temporal derivative of the opening near the fracture contour. This specific feature yields the 'duality' of the HF problem: being actually parabolic at the central part of the fracture, it is described by the equation of linear wave in a near-front zone.

(ii) When applied to numerical simulation of HF, the modified theory reveals three distinct issues important for developing efficient algorithms. These are: (a) *modeling the central part* of a HF; (b) *modeling the near-front zone*, and (c) *tracing* a propagating contour. For the *central part*, the analysis shows, as mentioned, that the system is actually parabolic; this feature leads to a highly stiff system of ODE after a spatial discretization. For the *near-front* zone, it is established that the system is hyperbolic: the zone propagates nearly as a simple wave. For *tracing a contour*, we state that it is a separate problem consisting of reconstruction the front. The latter may be easily performed on each or selected time steps of integration the SE by building the envelope to distance-circles or/and using marker particles.

(iii) After a spatial discretization, the three issues clearly appear in three distinct groups of boundary elements (internal, ribbon and tip elements), firstly distinguished in the ILSA. A small lag may be either neglected or easily accounted for by asymptotic matching suggested by the author. After neglecting the lag, fluid and elasticity equations refer to all the three groups, while in quite specific forms. It is highlighted that the fluid continuity equations, discretized by central differences, *identically* satisfy the global mass balance. We outline that this fact has a highly favourable effect on calculations. Namely, independently on whatever great errors in fluxes on common sides of neighbouring elements, the external fluid influx during any time interval *identically* equals to the change of the total fracture volume during this interval. This is of significance for assigning the fluxes on common sides of tip and ribbon elements.

(iv) For the *central part* of a HF, it is established that, despite highly restrictive CFL condition, *an acceptable time step* $\Delta t$ *is not prohibitively small*. This implies that explicit time stepping may be efficiently used even without acceleration(s), when the mesh size $\Delta x$ exceeds 0.05 of a typical half-size of the central part.

(v) For the *near-front zone*, its asymptotic analysis leads to the concept of the *universal asymptotic umbrella* (UAU), which universally connects the main characteristics of the near-front zone (the opening, the propagation speed, the distance to the front). The UAU provides a) the right hand sides of the discretized SE, and b) the fluxes on common sides of tip and ribbon elements. The discretized SE are assigned to centers of ribbon elements. Being separated, they may be efficiently integrated on large time steps of *implicit* schemes by employing specific features of the UAU. The fluxes on common sides of tip and ribbon elements may be assigned either by using the statistical approach by A. Stepanov, or by direct employing the UAU.

(vi) Another feature, important for numerical modeling, is that the propagation in the form of nearly simple wave occurs in a notable part of the zone adjacent to the front. For a homogeneous medium, the size of this zone is as large as 0.2 of a typical fracture half-length. This implies that the *CFL stability condition for a near-front zone is much less restrictive than for the central part* of the fracture. For a homogeneous medium, the size of the zone propagating as a simple wave is large enough to employ a rough mesh with some 5-10 elements along a typical half-length of the homogeneous region. Joined with the similar statement for the central part, this leads to the key conclusion that a rough spatial mesh may serve when modeling HF propagation in a homogeneous region. For an inhomogeneous medium with stress barriers, the CFL condition becomes even less restrictive due to drastic drop of the propagation speed. This suggests that a local fine mesh may be used in a near-front zone to model HF propagation after overcoming a barrier. However, the wave-like propagation of the near-front zone makes preferable upwind schemes of time stepping (especially, when employing *explicit* integration).

(vii) The results of points (iv) and (vi) mean that *explicit time stepping may be competitive and even superior* over implicit integration as regards to time expense. For *rough* spatial meshes, the time expense of explicit methods is on the level of implicit methods. For *fine* meshes on the central part of a HF, the restriction of the CFL condition is overcome by easily implemented acceleration(s) employing fast multipole methods (FMM), or fast Fourier transform (FFT), or/and parallel calculations.

(viii) With the shortcoming of excessive time expense removed, the computational advantages of explicit integrations become available. Specifically, in HF problems, *explicit integration involves merely matrix-to-vector multiplications with a fixed matrix*. The latter, although fully populated, is found in advance before temporal integration. In contrast with implicit methods, the explicit integration avoids: a) repeated building and inversion of ill-conditioned matrices on steps of fixed-point iterations for solving a system of strongly non-linear algebraic equations; b) the need in construction a proper preconditioner; and c) iterations for solving the linearized preconditioned system. Of special significance is that explicit integration is of immediate use for simulation HF driven by non-Newtonian fluid with arbitrary rheology.

(ix) Tracing *the front* presents a separate mostly geometrical problem. It includes a) *reconstruction* of the fracture contour on some steps of temporal integration, and b) *updating* the collections of internal, ribbon and tip elements. The reconstruction of the front consists of building the envelope to distance-circles centered at ribbon elements or/and employing marker particles. The need in updating the collections is easily checked on selected steps of the front reconstruction (when building the envelope), or in parallel with temporal integration (for marker particles). The updating itself does not involve computational difficulties.

(x) Numerical experiments for the bench-mark KGD and axisymmetric problems confirm the theoretical conclusions and estimations. They have also displayed high stability of explicit integration to extremely large (some orders of magnitude) perturbations of initial conditions. This is of special significance for practical applications, in which the initial conditions are quite uncertain.

(xi) Substitution of exact bench-mark values into the left and right hand sides of the spatially discretized equations displayed smoothing effect of temporal integration by both explicit and implicit methods. Despite of revealed inaccurate evaluation of the flux divergence on a time step, the openings, pressure and front positions, resulting after temporal integration, are quite accurate. The smoothing effect is essentially due to the outlined fact that the system of ODE used identically meets the mass conservation law on each time step and, consequently, on any time interval.

The conclusions suggest developing codes employing *explicit* integration as a perspective computational means for modeling HF. In author's view, it is reasonable to focus on fine meshes, needed, say, near local inhomogenuities. For fine meshes, the bottle-neck is the multiply repeated operation of matrix-to-vector multiplication. This operation may be drastically speeded up by using (i) fast multipole methods (for arbitrary meshes), (ii) fast Fourier transform (for uniform meshes), (iii) parallel computations.

**Appendix. Efficient solving speed equations on steps of implicit integration**

The system of ODE (26), (27) written in a vector form is $dz/dt = f(t,z)$, where z is a vector-column of unknowns (openings and distances to the front). Consider quite a general scheme of its integration: $\Delta z/\Delta t = (1-\omega)f(t, z(t)) + \omega f(t + \Delta t, z(t + \Delta t))$, where the parameter $\omega$ may be chosen in the interval $0 \leq \omega \leq 1$. The case $\omega = 0$ corresponds to the explicit forward Euler scheme. When $\omega \neq 0$, the scheme is implicit; in particular, $\omega = 1$ and $\omega = 1/2$ correspond, respectively, to the implicit backward Euler and trapezoid scheme. When applied to (26), (27), an implicit scheme requires solving strongly non-linear algebraic system in $w(t + \Delta t)$. The system is solved by fixed-point iterations. An iteration is performed with frozen arguments of the operator $\Lambda_f(w_f, p_f)$ in (26) and with frozen arguments $w_{Rj}$ in (27). With $r_j$ and $v_{*j}$ found from the ODE (27), the part (26) becomes linear algebraic system in $w_f$.

The SE equations (27) are separated. With the opening $w_{Rj}$ frozen, the speed $v_{*j} = v_*(t + \Delta t)$ and the distance $r_j = r(t + \Delta t)$ in each of them are easily found by writing the corresponding implicit scheme as

$$v_*(t + \Delta t) = \varphi_v(w_{Rj}, r(t + v_{*av}\Delta t)) \qquad (A1)$$

with $v_{*av} = (1-\omega)v_*(t) + \omega v_*(t + \Delta t)$. and $r(t + \Delta t) = r(t) + v_{*av}\Delta t$. In (A1), $r(t)$ and $v_*(t)$ are known at the beginning of the time step; $w_{Rj}$ is known as the frozen value of the opening at $t + \Delta t$, defined on an iteration step of solving the ODE (26).

The positive solution to (A1) is easily found as the intersection of non–negative functions $v_1 = v_*(t + \Delta t)$ and $v_2 = \varphi_v(w_{Rj}, r(t + v_{*av}\Delta t))$, representing, respectively, the left and right hand sides of (A1). When $v_*(t + \Delta t)$ grows from zero to infinity, the first of them linearly grows from zero to infinity, while the second decreases from the positive value $\varphi_0 = \varphi_v(w_{Rj}, r(t) +$ $(1 - \omega)v_*(t)\Delta t))$ to zero. Obviously, the solution is unique. The intersection point may be promptly found by a highly efficient subroutine. Finally, $r(t + \Delta t) = r(t) + v_{*av}\Delta t$.

Specifically, for the popular backward Euler scheme ($\omega = 1$), $v_{*av} = v_*(t + \Delta t)$. Therefore, $v_1 = v_*$, while $v_2 = \varphi_v(w, r(t) + v_*(t + \Delta t)\Delta t)$. Having the intersection of $v_1$ and $v_2$, the radius at $t + \Delta t$ is: $r(t + \Delta t) = r(t) + v_*(t + \Delta t)\Delta t$.

Since the SE is the inversion of the UAU (9) in the velocity, the pair $v_*(t + \Delta t)$, $r(t + \Delta t)$ satisfies equation (9) for the UAU, as well. Then it becomes evident that this pair also satisfies the inversion of the UAU (9) in the distance $r(t + \Delta t)$ to the front:

$$r(t + \Delta t) = \varphi_r(w_{Rj}, v_*(t + \Delta t)) \tag{A2}$$

with $v_*(t + \Delta t) = [r(t + \Delta t) - r(t)]/\Delta t$.

Again the unique positive solution to (A2) is easily and highly efficient found as the intersection of non-negative functions $r_1 = r(t + \Delta t)$ and $r_2 = \varphi_r(w_{Rj}, [r(t + \Delta t) - r(t)]/\Delta t)$. The first of them linearly grows from zero, while the second decreases from infinity at $r(t + \Delta t) = r(t)$ to zero when $r(t + \Delta t)$ goes to infinity.

In the implicit algorithm, suggested in [Peirce and Detournay 2008], the authors write and use the equation (A2) to find $-\tau = r(t + \Delta t)$ and $v_*(t + \Delta t)$ for three limiting regimes (toughness, viscosity and leak-off dominated) in the case of a Newtonian fluid, when the asymptotic umbrella has the monomial form (10). In these cases, the exponent $\alpha$ in (10) equals, respectively, 1/2, 2/3 and 5/8 for the three regimes considered, while the speed enters the right hand side of (10) as the factor $v_*{}^\beta$, where $\beta$ equals, respectively, 0, 1/3 and 1/8.

For the toughness dominated regime, the solution is $-\tau = -(w_{Rj}E'/K'_{IC})^2$. Tis solution is mentioned in Sec. 2 when discussing the SE (8) for the classical condition of linear fracture mechanics. The same result also follows when including arbitrary small amount of viscosity into the parameter entering the UAU for an intermediate regime between the toughness and viscosity dominated regimes (see [Linkov 2014 a, 2015]).

For the two remaining limiting regimes, equation (A2) becomes $\tau^{\gamma+1} - \tau_0\,\tau^\gamma - b = 0$, where $\gamma = \alpha/\beta$ and $b$ is a constant proportional to the product $w_{Rj}{}^{1/\beta}\Delta t$. The authors of the paper [Peirce and Detournay 2008] proved that this equation has unique positive solution. The discussion above proves that the positive solution exists and unique for any physically significant asymptotic umbrella (9). The given evidence is constructive; thus, it provides an easy means to efficiently find $r(t + \Delta t)$ and $v_*(t + \Delta t)$ for any such UAU (9).

Clearly, the UAU (9) itself may be used in the line discussed. For the Euler backward scheme, it is sufficient to substitute into it either $v_*(t + \Delta t) = [r(t + \Delta t) - r(t)]/\Delta t$, or $r(t + \Delta t) = r(t) + v_*\Delta t$. Then the left hand side of (9) is constant, while its right hand side is non-decreasing positive function of both arguments. The positive solution, defined by the intersection, is unique.

**Acknowledgments. T**he support of the National Science Centre of Poland (Project Number 2015/19/B/ST8/00712) is gratefully acknowledged. The author also appreciates the valuable assistance of MSc Alexey Stepanov in performing detailed calculations for the benchmark KGD and axisymmetric problems.

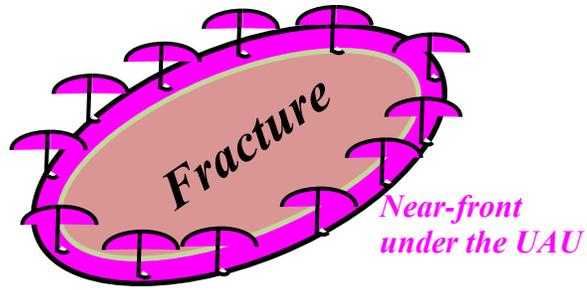

Fig . 1

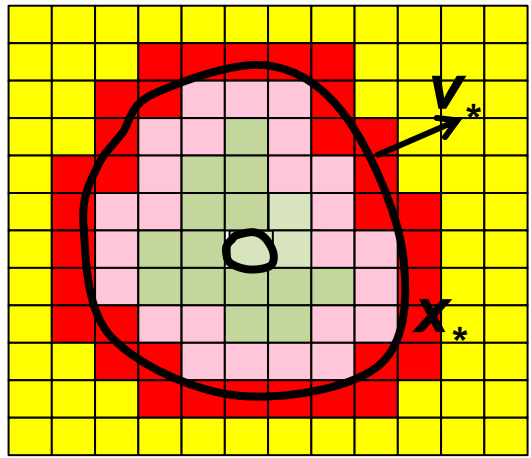

Fig . 2

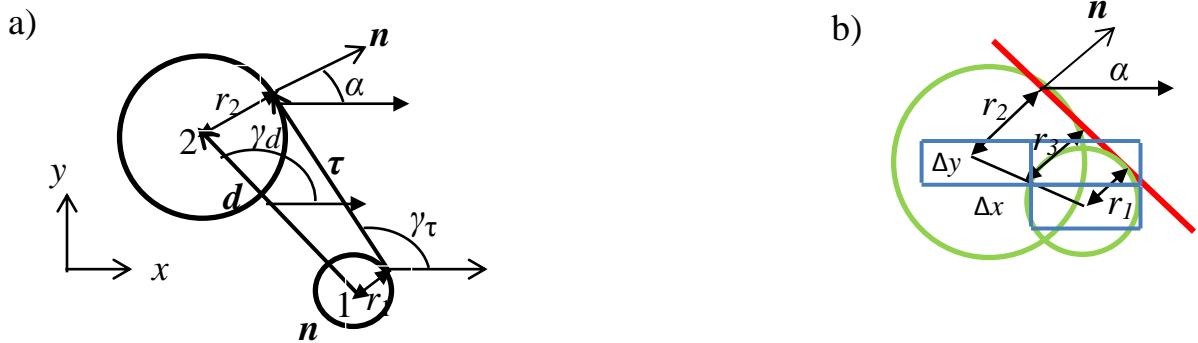

Fig. 3

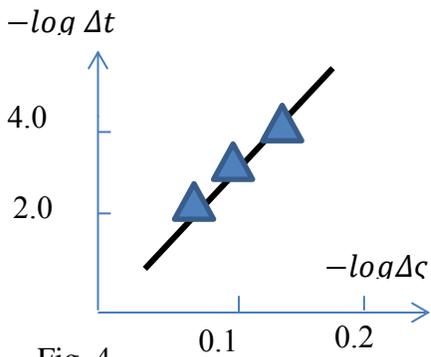

Fig. 4

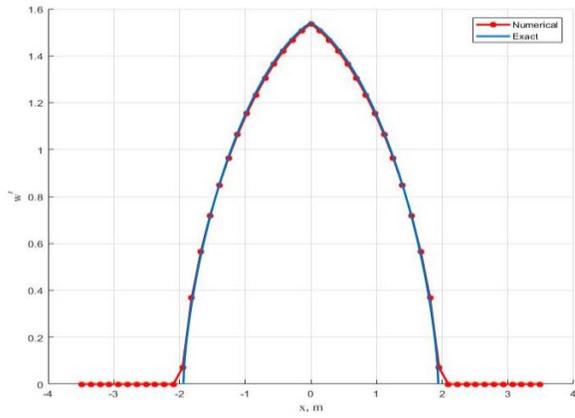

Fig. 5. Opening profiles for a Newtonian fluid ($n = 1$)

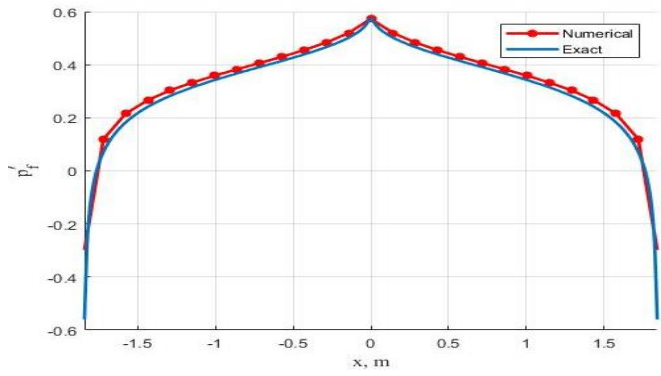

Fig. 6. Opening profiles for a non-Newtonian fluid ($n = 0.6$)